\documentclass[11pt]{article}
\usepackage{fullpage,proof,amssymb,amsmath,epsfig,stmaryrd}
\makeatletter
\newcommand{\singlespacing}{\let\CS=\@currsize\renewcommand{\baselinestretch}{1}\small\CS}
\newcommand{\doublespacing}{\let\CS=\@currsize\renewcommand{\baselinestretch}{1.75}\small\CS}
\newcommand{\normalspacing}{\let\CS=\@currsize\renewcommand{\baselinestretch}{\BLS}\small\CS}
\makeatother

\newtheorem{thm}{Theorem}[section]

\newtheorem{prop}[thm]{Proposition}
\newtheorem{rem}[thm]{Remark}

\newtheorem{defin}[thm]{Definition}

\newcommand{\rarr}{\rightarrow}
\newcommand{\lrarr}{\longrightarrow}

\newcommand{\ox}{\otimes}

\newcommand{\ent}{\vdash}

\newcommand{\girpar}{\bindnasrepma}

 \def\pushright#1{{
    \parfillskip=0pt            
    \widowpenalty=10000         
    \displaywidowpenalty=10000  
    \finalhyphendemerits=0      
    \leavevmode                 
    \unskip                     
    \nobreak                    
    \hfil                       
    \penalty50                  
    \hskip.2em                  
    \null                       
    \hfill                      
    {#1}                        
    \par}}                      
 \def\qed{\pushright{\rule{2mm}{3mm}}\penalty-700 \smallskip}

\newenvironment{proof}[1]{\begin{trivlist} \item[{\bf ~Proof} #1.]}%
{\qed\end{trivlist}}

\makeatletter

\newdimen\w@dth

\def\setw@dth#1#2{\setbox\z@\hbox{\scriptsize $#1$}\w@dth=\wd\z@
\setbox\@ne\hbox{\scriptsize $#2$}\ifnum\w@dth<\wd\@ne \w@dth=\wd\@ne \fi
\advance\w@dth by 1.2em}

\def\t@^#1_#2{\allowbreak\def\n@one{#1}\def\n@two{#2}\mathrel
{\setw@dth{#1}{#2}
\mathop{\hbox to \w@dth{\rightarrowfill}}\limits
\ifx\n@one\empty\else ^{\box\z@}\fi
\ifx\n@two\empty\else _{\box\@ne}\fi}}
\def\t@@^#1{\@ifnextchar_ {\t@^{#1}}{\t@^{#1}_{}}}

\def\t@left^#1_#2{\def\n@one{#1}\def\n@two{#2}\mathrel{\setw@dth{#1}{#2}
\mathop{\hbox to \w@dth{\leftarrowfill}}\limits
\ifx\n@one\empty\else ^{\box\z@}\fi
\ifx\n@two\empty\else _{\box\@ne}\fi}}
\def\t@@left^#1{\@ifnextchar_ {\t@left^{#1}}{\t@left^{#1}_{}}}

\def\two@^#1_#2{\def\n@one{#1}\def\n@two{#2}\mathrel{\setw@dth{#1}{#2}
\mathop{\vcenter{\hbox to \w@dth{\rightarrowfill}\kern-1.7ex
                 \hbox to \w@dth{\rightarrowfill}}%
       }\limits
\ifx\n@one\empty\else ^{\box\z@}\fi
\ifx\n@two\empty\else _{\box\@ne}\fi}}
\def\tw@@^#1{\@ifnextchar_ {\two@^{#1}}{\two@^{#1}_{}}}

\def\tofr@^#1_#2{\def\n@one{#1}\def\n@two{#2}\mathrel{\setw@dth{#1}{#2}
\mathop{\vcenter{\hbox to \w@dth{\rightarrowfill}\kern-1.7ex
                 \hbox to \w@dth{\leftarrowfill}}%
       }\limits
\ifx\n@one\empty\else ^{\box\z@}\fi
\ifx\n@two\empty\else _{\box\@ne}\fi}}
\def\t@fr@^#1{\@ifnextchar_ {\tofr@^{#1}}{\tofr@^{#1}_{}}}

\newdimen\W@dth
\def\setW@dth#1#2{\setbox\z@\hbox{$#1$}\W@dth=\wd\z@
\setbox\@ne\hbox{$#2$}\ifnum\W@dth<\wd\@ne \W@dth=\wd\@ne \fi
\advance\W@dth by 1.2em}

\def\T@^#1_#2{\allowbreak\def\N@one{#1}\def\N@two{#2}\mathrel
{\setW@dth{#1}{#2}
\mathop{\hbox to \W@dth{\rightarrowfill}}\limits
\ifx\N@one\empty\else ^{\box\z@}\fi
\ifx\N@two\empty\else _{\box\@ne}\fi}}
\def\T@@^#1{\@ifnextchar_ {\T@^{#1}}{\T@^{#1}_{}}}

\def\T@left^#1_#2{\def\N@one{#1}\def\N@two{#2}\mathrel{\setW@dth{#1}{#2}
\mathop{\hbox to \W@dth{\leftarrowfill}}\limits
\ifx\N@one\empty\else ^{\box\z@}\fi
\ifx\N@two\empty\else _{\box\@ne}\fi}}
\def\T@@left^#1{\@ifnextchar_ {\T@left^{#1}}{\T@left^{#1}_{}}}

\def\Tofr@^#1_#2{\def\N@one{#1}\def\N@two{#2}\mathrel{\setW@dth{#1}{#2}
\mathop{\vcenter{\hbox to \W@dth{\rightarrowfill}\kern-1.7ex
                 \hbox to \W@dth{\leftarrowfill}}%
       }\limits
\ifx\N@one\empty\else ^{\box\z@}\fi
\ifx\N@two\empty\else _{\box\@ne}\fi}}
\def\T@fr@^#1{\@ifnextchar_ {\Tofr@^{#1}}{\Tofr@^{#1}_{}}}

\def\Two@^#1_#2{\def\N@one{#1}\def\N@two{#2}\mathrel{\setW@dth{#1}{#2}
\mathop{\vcenter{\hbox to \W@dth{\rightarrowfill}\kern-1.7ex
                 \hbox to \W@dth{\rightarrowfill}}%
       }\limits
\ifx\N@one\empty\else ^{\box\z@}\fi
\ifx\N@two\empty\else _{\box\@ne}\fi}}
\def\Tw@@^#1{\@ifnextchar_ {\Two@^{#1}}{\Two@^{#1}_{}}}

\def\to{\@ifnextchar^ {\t@@}{\t@@^{}}}
\def\from{\@ifnextchar^ {\t@@left}{\t@@left^{}}}
\def\two{\@ifnextchar^ {\tw@@}{\tw@@^{}}}
\def\tofro{\@ifnextchar^ {\t@fr@}{\t@fr@^{}}}
\def\To{\@ifnextchar^ {\T@@}{\T@@^{}}}
\def\From{\@ifnextchar^ {\T@@left}{\T@@left^{}}}
\def\Two{\@ifnextchar^ {\Tw@@}{\Tw@@^{}}}
\def\Tofro{\@ifnextchar^ {\T@fr@}{\T@fr@^{}}}

\makeatother

\newcommand{\cO}{\mathcal{O}}
\newcommand{\cP}{\mathcal{P}}

\newcommand{\hi}{{\cal H}}
\newcommand{\ind}{\mathit {End}}
\newcommand{\psau}{\psi_1^{\uparrow}}
\newcommand{\psad}{\psi_1^{\downarrow}}
\newcommand{\psbu}{\psi_2^{\uparrow}}
\newcommand{\psbd}{\psi_2^{\downarrow}}
\newcommand{\psauk}{|\psi_1^{\uparrow}\rangle}
\newcommand{\psadb}{\langle\psi_1^{\downarrow}|}
\newcommand{\psadk}{|\psi_1^{\downarrow}\rangle}
\newcommand{\psaub}{\langle\psi_1^{\uparrow}|}
\newcommand{\psbuk}{|\psi_2^{\uparrow}\rangle}

\newcommand{\psbub}{\langle\psi_2^{\uparrow}|}

\newcommand{\ket}[1]{| #1 \rangle}

\newcommand{\proj}[1]{| #1 \rangle\langle #1 |}

\begin{document}

\title{Discrete Quantum Causal Dynamics}

\author{
\begin{tabular}[t]{c}
        Richard F. Blute\thanks{
Research supported in part 
by NSERC.
}\\Ivan T. Ivanov\\
        {\small Department of Mathematics}\\
        {\small \rule{0mm}{3mm} and Statistics}\\
        {\small University of Ottawa}\\
        {\small Ottawa, Ontario, Canada}\\
        \texttt{rblute,iti@mathstat.uottawa.ca}\\ 
\end{tabular}
\and
\begin{tabular}[t]{c}
        Prakash Panangaden\footnotemark[1]\\ 
        {\small School of Computer Science}\\
        {\small McGill University}\\
        {\small Montr\'eal, Qu\'ebec, Canada}\\
        \texttt{prakash@cs.mcgill.ca}
    \end{tabular}
}

\maketitle

\renewcommand{\girpar}{\bindnasrepma}

\begin{abstract}  We give a mathematical framework to
describe the evolution of an open quantum systems subjected
to finitely many interactions with classical apparatuses.  The
systems in question may be composed of distinct, spatially
separated subsystems which evolve independently but may also
interact. This evolution, driven both by unitary operators
and measurements, is coded in a precise mathematical
structure in such a way that the crucial properties of
causality, covariance and entanglement are faithfully
represented.  We show how our framework may be expressed
using the language of (poly)categories and functors. 
Remarkably, important physical consequences - such as
covariance - follow directly from the functoriality of our
axioms.

We establish strong links between the physical
picture we propose and linear logic. Specifically we show
that the refined logical connectives of linear logic can be
used to describe the entanglements of subsystems in a
precise way. Furthermore, we show that there is a precise 
correspondence between the evolution of a given 
system and deductions in a certain formal logical system
based on the rules of linear logic.

This framework generalizes and enriches both
causal posets and the histories approach to quantum
mechanics.
\end{abstract}

\section{Introduction}  

We propose a uniform scheme for describing a quantum system,
interacting with a network of classical objects.  The
system in question may be composed of distinct spatially
separated subsystems which evolve independently, but may
also interact with each other at various points as well as
with the classical objects.  When analyzing physical
laboratory experiments on quantum systems, we frequently
abstract away from the concrete experimental setup and from
the particular details of the machinery involved.  What we
usually keep is the description of the quantum system - and
its spatially separated subsystems - in terms of wave
functions or density matrices and unitary operators as well
as the changes of the quantum system induced by the
interactions with classical devices. Crucial properties of
the evolution such as the causal ordering, covariance of the
description for different observers and quantum entanglement
between distinct subsystems should be completely reflected
in any such description.

The basis of our representation is the graph of events and
causal links between them.  An event could be one of the
following: a unitary evolution of some subsystem, an
interaction of a subsystem with a classical device (a
measurement) or perhaps just the coming together or
splitting apart of several spatially separated  subsystems. 
Events will be depicted as vertices of a directed graph.  The
edges of the graph will represent the causal relations
between the different events.  The vertices of the graph are
then naturally labelled with operators representing the
corresponding processes.

Of course, the processes of unitary evolution and
measurement take a certain amount of time; but we are only
interested in the causal relations between such events and
this allows us to consider them as point-like vertices on the
graph.  Thus we are thinking of the duration between events
as being longer than the duration of an event so that no
causal information is lost when we represent interactions as
events.

The structure described thus far reflects the kinematical
properties of the quantum system.  To describe the dynamics
we need a composition of the operators assigned to the
vertices of the graph.  This composition is most conveniently
described in terms of a composition in a specific
mathematical structure, namely a polycategory generated by
the graph.  The whole description could then be concisely
summarized by noticing that we have a functor from this
polycategory to the polycategory of Hilbert spaces.   This
functor captures the dynamics of the system.

Causal relations are made explicit and we prove that no
influences breaking causality arise in our scheme.  The
possible entanglement between spatially separated subsystems
- represented by distinct edges of the graph - is also
accounted for.  Thus, our framework allows one to represent
locality of interaction - i.e.\ causal influences do not
propagate outside the causal ``cone'' - while allowing the
expression of nonlocal correlations which occur when one has
quantum entanglement.  The tension between causal evolution
and quantum entanglement is resolved.

The categorical framework that we use is intimately
connected with linear logic. Linear logic was
originally introduced~\cite{Girard87} as a logic intended
for a finer analysis of the way resources are consumed
during the course of a proof.  This logic has had a
significant impact on the theory of computation as well as
such far-flung areas as linguistics and pure mathematics. 
In the present paper, the connectives of linear logic will
be used to express the existence or nonexistence of nonlocal
correlations. What we will introduce is a deductive system
based on the graph-theoretic structure of the system
that precisely picks out the spatial slices of physical
interest. Thus evolution of the system corresponds to 
logical deductions within this deductive system.
For an expository introduction
to linear logic, see the review by Girard~\cite{Girard95} or
the brief exposition in the appendix.

\subsection{Relation to other work}

Next we outline the relations of our proposal to some
recent approaches to quantum mechanics and quantum gravity.

\subsubsection{Consistent and decoherent histories}

The \emph{consistent histories} approach to quantum
mechanics due to Griffiths and Omn\`es
\cite{Griffiths96,Omnes94} was formulated with the aim of
shedding new light on the conceptual difficulties of the
theory.  A closely related proposal with different
motivation is the \emph{decoherent histories} approach to
quantum cosmology of Gell-Mann and Hartle
\cite{Gell-Mann93}.  The basic ingredient in both approaches
is the notion of a \emph{history} of the quantum system
described by a sequence of projection operators in the
Hilbert space of the system, for a succession of times.  The
goal of quantum mechanics is to determine the probability of
an event or a sequence of events, thus one might hope to
assign probabilities to the histories of the quantum
system.  In order for the probabilities to be additive in
the usual sense, the histories have to be mutually
noninterfering.  Sets of histories obeying this condition
are selected with the use of a special bilinear form on
histories - the decoherence functional.  

A particular history is mathematically represented as a
linearly ordered sequence of projection operators in the
Hilbert space of the quantum mechanical system.  But the
linear causal ordering of the events in a history is too
restrictive in many experimental situations, in particular
when analyzing spatially separated entangled quantum
systems.  This issue is even more pressing for quantum
cosmology considerations.  An application of the histories
approach to quantum field theory on a curved space-time
\cite{Blencowe91} must assume the existence of a globally
hyperbolic manifold, and thus via the associated foliation,
a linear ordering of the histories of the quantum field.

Our proposal for describing the evolution of an open quantum
system can be considered as describing a single history in a
set of histories.  The important point is that events are no
longer linearly ordered by temporal order but, rather,
partially ordered with respect to the causal order.  This
allows one to capture the notion of causal evolution in a
manifestly covariant fashion.  The consistency/decoherence
condition for histories has an immediate generalization for
histories described by more general graphs as proposed here.

\subsubsection{Causal sets}

\emph{Causal sets} form the basis of an approach to quantum
gravity mainly advocated by R. Sorkin and collaborators
\cite{Bombelli87,Sorkin91}, where the basic idea is to take
the notion of causality as the primitive. In classical
relativity, the structure of the space-time manifold
together with a metric of Lorentzian signature determines
the causality relation.  An important observation is that
the causal structure is conformally invariant, i.e.\
determined by only the conformal equivalence class of the
metric and hence more primitive than the metric.  Various
proposals for quantum gravity - for example, the twistor
program~\cite{Penrose72} - have taken as their point of
departure the idea that the causal structure is more
fundamental than the metric structure.
  
In the causal sets approach, one takes the point of view
that, at the smallest length scales, spacetime is inherently
discrete and that the causal structure, the ``light cones'',
are fundamental.  This leads naturally to the idea of a
partially ordered set (poset for short) where the elements
are events and two events are related by causality.  The
main interest is in approximating continuous spacetimes with
such structures and defining processes that would generate
these structures, with a view to an eventual theory of
quantum gravity.  Though the aims are rather different the
issues connected with causality are closely related.

Causal sets are further motivated by the idea that a
discrete structure would avoid the singularities that plague
physics (both classical and quantum).  The assumption that
space-time should be a continuous manifold is one of the
ingredients that leads to the problematic singularities of
quantum field theory and general relativity.  In the causal
sets approach, space-time is a discrete structure, thus
possibly avoiding these singularities, the idea being that
at the Planck scale, continuous geometry gives way to
discrete geometry.  

One way to think of this is that one approximates a manifold
as one ``sprinkles'' more and more points into the causal
set in a uniform fashion.  More formally, one would want to
obtain a manifold as the categorical limit of a diagram of
posets and embeddings \cite{Maclane98}. Applications and
extensions of these ideas can be found in papers such as
\cite{Markopoulou00,Markopoulou97,Raptis00b}, although this
list is by no means exhaustive.  In our approach we are not
thinking about generating the spacetime through such
limiting processes, but the idea of a causal set is implicit
in our work.  For us, a finite causal set is the kinematical
framework on which we describe evolution and information
flow.

\subsubsection{Quantum causal histories}

The notion of \emph{quantum causal history} was introduced
by Markopoulou in \cite{Markopoulou00}.  One begins with
a poset (causal set) and assigns Hilbert spaces to the
vertices and evolution operators to sets of edges.  The
assignment must satisfy properties analogous to
functoriality.  However, within this framework, one is
quickly led to violations of causality - as the author
herself notes - essentially because the slices used are
``too global.''  She mentions the possibility of working
with a dual view.  In fact, in our work, we take such a
dualized view as our starting point.  In other words we
assign operators representing evolution or measurement to
vertices and Hilbert spaces to the edges, in a way
satisfying (poly)functoriality.

\subsection{The Importance of Categories} A category can be
seen as a generalization of a poset in the following sense. 
A poset merely records that an element $x$ is less than $y$
but a category keeps track of the different ways in which
$x$ might be less than
$y$.  For example, in logic one might consider formulas
(denoted by Greek letters like $\phi$, $\psi$ etc.\ ) and
the relationship of provability between them.  Thus one
would write $\phi\vdash\psi$ to mean that starting from the
assumption $\phi$ one can prove $\psi$.  This gives rise to
a transitive and reflexive relation; if one considers
equivalence classes of formulas (two formulas being
equivalent if each can be used to prove the other) we get a
poset.  However, if we are interested in distinguishing
distinct proofs we need to keep track of the different ways
in which $\phi$ can be used to prove $\psi$.  Thus formulas
as objects and proofs as morphisms can be organized into a
category.  

In a poset when one writes $x\leq y$ then, depending on the
context, one is stating something like the following:
\begin{itemize}
\item $x$ is less than $y$;
\item $x$ precedes $y$;
\item $x$ implies $y$.
\end{itemize}
\noindent or any of several other possibilities.  In a
causal set, we have in mind that $x$ causally precedes $y$.  

In the present work, we are particularly interested in
modelling the idea that information can flow from one event
to another in a number of different ways, \emph{along
different paths or channels}.  We would like to keep track
of all these various independent paths.  The structure of a
poset is inadequate for achieving this, as we would like to
say that $x\leq y$ in several different ways.  This
naturally suggests that we pass from posets to more general
graphs and eventually to categories.

Many recent experiments feature spatially distributed
quantum systems. When entangled quantum subsystems come back
together in the same spacetime region, the description of the
resulting system is causally influenced by all events in the
paths of the subsystems.  In particular a past event could
influence the future events in several distinct ways through
different paths.  Our scheme is well adapted for analyzing
experiments featuring spatially separated quantum entangled
entities and could be used in the field of quantum
information processing to analyze information flow
situations.

\subsection{Contents of the present paper}

Section~\ref{causal} presents the basic ideas of our scheme
via an example. Section~\ref{dyn} discusses the basic
physical ideas involved.  In the first subsection we review
the notions of measurements and interventions. In the next
subsection we give the dynamical prescription in a special
case and in the final subsections we give the general
prescription and prove covariance.  In section~\ref{poly} we
review basic facts about polycategories and their
construction.  We also describe the polycategory of Hilbert spaces 
and intervention operators 
we will be using. In section~\ref{logic} we give a logical
presentation of polycategories and establish the connection
between our structures and linear logic. A functorial version of 
our dynamical prescription is then presented.   
We end with a discussion on further
applications of our scheme.  
It is our hope that this paper will
interest members of several different communities within
mathematics, logic and physics.
 
\section{Causal information flow via
examples}\label{causal}  

Consider a quantum system evolving in space-time while being
subjected to interactions with classical observers at a
number of points.  The causal and spatio-temporal relations
in the system will be represented by a directed acyclic
graph (hereafter called a \emph{dag}).  The vertices of the
graph - which will be drawn as boxes - represent the events
in the evolution of the system.  An event could be a
measurement by a classical observer, a local unitary
evolution or just a splitting of a subsystem into several
spatially separated subsystems, which however could still
share an entangled common state.  The propagation of the
different subsystems will be indicated by the edges of the
graph.

There are a number of causal relations between edges and
vertices.  A vertex $v_1$ is said to \emph{immediately
precede} $v_2$ if there is a (directed) edge from $v_1$ to
$v_2$.  We write $v_1 \leq v_2$ for the reflexive transitive
closure of immediate precedence; thus $v \leq v$ always
holds and $v_1 \leq v_2$ means that there is a
\emph{directed path} from $v_1$ to $v_2$ (possibly of length
zero).  When $v_1 \leq v_2$ we sometimes say $v_1$ is ``to
the past of'' $v_2$ and dually ``$v_2$ is to the future of
$v_1$.''  When we draw a poset we typically leave out the
self-loops and only draw the minimal number of edges needed
to infer all the others; the so-called ``Hasse diagram'' of
the poset. We note that  our graphs will have initial and
final  ``half-edges'', i.e. edges with only one endpoint. 
Physically we have some quantum states incoming (or
``prepared'') followed by some interactions and some
outgoing state.

The relation between vertices induces a causal relation
between edges.  We say that an edge $e_1$ is to the past of
another edge $e_2$ if the terminal vertex of $e_1$, say
$v_1$ and the initial vertex of $e_2$, say $v_2$, satisfy
$v_1 \leq v_2$.  Note that we could have $v_1 = v_2$.  An
initial edge is not to the future of any edge, nor is a
final edge to the past of any other edge.  If two edges are
not causally related, we say that they are ``spacelike
separated'' or acausal.  Note that two spacelike separated
edges could share a common terminal vertex or a common
initial vertex, (but since we have a graph, not both).  A
\emph{space-like slice} is defined as a set of pairwise
acausal edges.  Henceforth, whenever we say ``slice'' we
will always mean ``spacelike slice.''  Note that the initial
(or final) edges form a spacelike slice.  We call this the
\emph{initial (final) slice}.

For example for the graph of Figure~\ref{figureN} the set of
edges $\{e_c, e_d, e_e\}$ form a space-like slice.  Another
example is the set $\{e_f, e_d, e_e\}$.  The edges $e_a$ and
$e_b$ form the initial slice.   The edges $e_a,e_b,e_f$ and
$e_g$ are half-edges, with $e_a$ and $e_b$ initial, and
$e_f$ and $e_g$ final.

\begin{figure}[htb]
\begin{center}
\setlength{\unitlength}{3947sp}%
\begingroup\makeatletter\ifx\SetFigFont\undefined%
\gdef\SetFigFont#1#2#3#4#5{%
  \reset@font\fontsize{#1}{#2pt}%
  \fontfamily{#3}\fontseries{#4}\fontshape{#5}%
  \selectfont}%
\fi\endgroup%
\begin{picture}(1524,2124)(4789,-3223)
\thinlines
\special{ps: gsave 0 0 0 setrgbcolor}\put(4801,-1861){\framebox(300,300){$v_3$}}
\special{ps: gsave 0 0 0 setrgbcolor}\put(6001,-1861){\framebox(300,300){$v_4$}}
\special{ps: gsave 0 0 0 setrgbcolor}\put(4801,-2761){\framebox(300,300){$v_1$}}
\special{ps: gsave 0 0 0 setrgbcolor}\put(6001,-2761){\framebox(300,300){$v_2$}}
\special{ps: gsave 0 0 0 setrgbcolor}\put(4951,-2461){\vector( 0, 1){375}}
\special{ps: grestore}\special{ps: gsave 0 0 0 setrgbcolor}\put(4951,-2161){\line( 0, 1){300}}
\put(4731,-2161){$e_c$} 
\special{ps: grestore}\special{ps: gsave 0 0 0 setrgbcolor}\put(6151,-2461){\vector( 0, 1){375}}
\special{ps: grestore}\special{ps: gsave 0 0 0 setrgbcolor}\put(6151,-2161){\line( 0, 1){300}}
\put(6221,-2161){$e_e$}
\special{ps: grestore}\special{ps: gsave 0 0 0 setrgbcolor}\put(4951,-2461){\vector( 2, 1){600}}
\special{ps: grestore}\special{ps: gsave 0 0 0 setrgbcolor}\put(5551,-2161){\line( 2, 1){600}}
\put(5451,-2061){$e_d$}
\special{ps: grestore}\special{ps: gsave 0 0 0 setrgbcolor}\put(4951,-1561){\vector( 0, 1){375}}
\special{ps: grestore}\special{ps: gsave 0 0 0 setrgbcolor}\put(4951,-1186){\line( 0, 1){ 75}}
\put(4731,-1256){$e_f$}
\special{ps: grestore}\special{ps: gsave 0 0 0 setrgbcolor}\put(6151,-1561){\vector( 0, 1){375}}
\special{ps: grestore}\special{ps: gsave 0 0 0 setrgbcolor}\put(6151,-1186){\line( 0, 1){ 75}}
\put(6221,-1256){$e_g$}
\special{ps: grestore}\special{ps: gsave 0 0 0 setrgbcolor}\put(4951,-3211){\vector( 0, 1){225}}
\special{ps: grestore}\special{ps: gsave 0 0 0 setrgbcolor}\put(4951,-3061){\line( 0, 1){300}}
\put(4731,-3061){$e_a$}
\special{ps: grestore}\special{ps: gsave 0 0 0 setrgbcolor}\put(6151,-3211){\vector( 0, 1){225}}
\special{ps: grestore}\special{ps: gsave 0 0 0 setrgbcolor}\put(6151,-3061){\line( 0, 1){300}}
\put(6221,-3061){$e_b$}
\special{ps: grestore}
\end{picture}
\end{center}
\caption{}
\label{figureN}
\end{figure}

Associated with any edge $e_i$ is an observer who has access
to a subsystem of the complete quantum system. Thus the
edges represent local information. Each edge $e_i$ is
assigned a density matrix $\rho_i$ in a Hilbert space
$\hi_i$\footnote{Throughout the paper, we assume that the
graph and the dimensions of all Hilbert spaces are
finite.}.  The density matrix $\rho_i$ describes the
knowledge about the quantum system available to the local
observer at the edge $e_i$.  More generally density matrices
will be associated to space-like slices.  For a space-like
slice consisting of edges $\{e_{i_1}, \dots e_{i_p}\}$, the
assigned density matrix will be denoted $\rho_{i_1, \dots
i_p}$.  This density matrix describes the subsystem of the
whole quantum system for that space-like slice.  Every
space-like slice has also a Hilbert space which is the
tensor product of the Hilbert spaces of the edges forming
the slice.  However the density matrix associated with the
slice is not in general a tensor product of the density
matrices on the edges. If it were, we could not capture
non-local quantum correlations.

The graph of Figure~\ref{figureN}, represents a quantum
system $\mathit{Q}$ which starts evolving from a state in
which $\mathit{Q}$ consists of two spatially separated
subsystems $\mathit{Q}_a$ and $\mathit{Q}_b$ described by
density matrices $\rho_a$ and $\rho_b$ respectively, in
Hilbert spaces
${\hi}_a$ and ${\hi}_b$.  The initial edges $e_a$ and $e_b$
form the initial slice in this simple system.  We will
follow the convention that if the initial slice consists of
several edges, the initial state of the whole system is a
tensor product state, i.e.  the subsystems are not
entangled. For the above example, $\psi_{init} = \psi_a
\otimes \psi_b$ and $\rho_a =
\proj{\psi_a}$ and $\rho_b = \proj{\psi_b}$. Entangled
subsystems on distinct edges will always have at least one
event in the common past.  Thus we always explicitly
represent the interaction which caused the entanglement.

Each vertex $v_i$ of the graph is labelled with an operator
$T_i$ which describes the process taking place at the
corresponding event.  The operator $T_i$ at a given event
$v_i$ takes density matrices on the tensor product of
Hilbert spaces living on the incoming edges at $v_i$ to
density matrices on the tensor product Hilbert space of
outgoing edges.  The process at a vertex could be an
\emph{intervention}\footnote{Interventions are generalized
measurements where a quantum subsystem could be
discarded~\cite{Peres00}.  This will be discussed more fully
below.} corresponding to a positive operator-valued measure
(POVM)~\cite{Nielsen00,Peres95} or a unitary
transformation.  Or instead of an external or unitary action
there could be several quantum subsystems that come together
and then split apart, possibly in a different way.  We will
consider this last case as a particular instance of a
unitary evolution with identity evolution operator. As a
simple example, in the case of an event corresponding to
unitary evolution by a unitary operator
$U$, we have the usual expression:

\begin{equation}\label{inop}
\rho^{in} \ \mapsto \ \rho^{out} \ = \
U \rho^{in} U^\dagger
\end{equation}  

The general expression for an operator associated to an
event will be discussed fully in the next section, see
equation (\ref{inap}).

Here we will discuss some of the conditions such a dynamical
scheme has to satisfy in order to reflect causality and
other physical properties of the quantum system.  Causality
is the condition that the density matrix on a given edge
should not depend on the actions performed at vertices which
are acausal to this edge or are in its future.  For example,
referring back to Figure~\ref{figureN}, we would like any
quantum evolution rule to say that the density matrix at
$e_g$ is unaffected by the intervention at $v_3$ or the
density matrix at $e_f$ is unaffected by the intervention at
$v_2$.  A general unitary evolution between the states of
two space-like slices is easily shown to violate this
condition.  Therefore we need to incorporate some sort of
locality condition into the evolution scheme.

It is not hard to formulate such an evolution scheme.  For
example, one could work with the dual picture and have
evolution occur along edges with density matrices at the
vertices.  It is not hard to formulate rules which would
enforce causality properly in such a framework. 
Unfortunately this rules out quantum correlations across
spatially separated subsystems. Thus, the evolution scheme
cannot be too local because entangled subsystems of the
quantum system could fly apart and later come together at a
vertex.

Consider the system shown in Figure~\ref{figureD}.
\begin{figure}[htb]
\begin{center}

\setlength{\unitlength}{3947sp}%
\begingroup\makeatletter\ifx\SetFigFont\undefined%
\gdef\SetFigFont#1#2#3#4#5{%
  \reset@font\fontsize{#1}{#2pt}%
  \fontfamily{#3}\fontseries{#4}\fontshape{#5}%
  \selectfont}%
\fi\endgroup%
\begin{picture}(2724,3849)(3589,-6073)
\thinlines
\special{ps: gsave 0 0 0 setrgbcolor}
\put(4801,-3061){\framebox(300,300){$1_{{\hi}_{f}}$}}
\special{ps: gsave 0 0 0 setrgbcolor}
\put(3601,-4261){\framebox(300,300){$T_2$}}
\special{ps: gsave 0 0 0 setrgbcolor}
\put(6001,-4261){\framebox(300,300){$T_3$}}
\special{ps: gsave 0 0 0 setrgbcolor}
\put(4801,-5461){\framebox(300,300){$1_{{\hi}_{a}}$}}
\special{ps: gsave 0 0 0 setrgbcolor}\put(3751,-3961){\vector( 4, 3){600}}
\special{ps: grestore}\special{ps: gsave 0 0 0 setrgbcolor}\put(4351,-3511){\line( 4, 3){600}}
\put(4441,-3611){$\rho_d$}
\special{ps: grestore}\special{ps: gsave 0 0 0 setrgbcolor}\put(6151,-3961){\vector(-4, 3){600}}
\special{ps: grestore}\special{ps: gsave 0 0 0 setrgbcolor}\put(5551,-3511){\line(-4, 3){600}}
\put(5351,-3611){$\rho_e$}
\special{ps: grestore}\special{ps: gsave 0 0 0 setrgbcolor}\put(4951,-5161){\vector(-4, 3){600}}
\special{ps: grestore}\special{ps: gsave 0 0 0 setrgbcolor}\put(4351,-4711){\line(-4, 3){600}}
\put(4441,-4651){$\rho_b$}
\special{ps: grestore}\special{ps: gsave 0 0 0 setrgbcolor}\put(4951,-2761){\vector( 0, 1){450}}
\special{ps: grestore}\special{ps: gsave 0 0 0 setrgbcolor}\put(4951,-2386){\line( 0, 1){150}}
\put(4701,-2416){$\rho_f$}
\special{ps: grestore}\special{ps: gsave 0 0 0 setrgbcolor}\put(4951,-5886){\vector( 0, 1){125}}
\special{ps: grestore}\special{ps: gsave 0 0 0 setrgbcolor}\put(4951,-5836){\line( 0, 1){375}}
\put(4701,-5836){$\rho_a$}
\special{ps: grestore}\special{ps: gsave 0 0 0 setrgbcolor}\put(4951,-5886){\line( 0,-1){ 75}}
\special{ps: grestore}\special{ps: gsave 0 0 0 setrgbcolor}\put(4951,-5161){\vector( 4, 3){600}}
\special{ps: grestore}\special{ps: gsave 0 0 0 setrgbcolor}\put(5551,-4711){\line( 4, 3){600}}
\put(5351,-4651){$\rho_c$}
\special{ps: grestore}\end{picture}
\end{center}
\caption{}
\label{figureD}
\end{figure} The quantum system represented in this graph is
as follows.  The system is prepared in a state $\psi_a$ as
indicated by the density matrix $\rho_a =
\proj{\psi_a}$ on the incoming edge.  At the vertex $v_1$
the system splits into two spatially separated subsystems on
the edges $e_b$ and $e_c$ which, in general, are still
described by a global entangled state.  The local
transformations $T_2$ and $T_3$ will, in general, preserve
the entanglement and the global state will be still
entangled on the space-like slice
$\{e_d, e_e\}$.  The two subsystems come together at the
vertex $v_4$.  The two local density matrices $\rho_d$ and
$\rho_e$ are not sufficient to reconstruct the entangled
state of the system described by $\rho_f$.  The off-diagonal
terms of $\rho_f$ are not reflected in the local density
matrices, $\rho_d$ and $\rho_e$.  We need to include
information about the history of the state on the space-like
slice $\{e_d, e_e\}$ in order to reconstruct the global
state.  One possibility is to work with global space-like
slices, and show that the scheme is generally covariant in
the sense of being slice-independent.  In our functorial
approach, certain preferred (not necessarily global)
spacelike slices account for all entanglement.

The rules for constructing and labeling the graphs given so
far reflect the kinematics of the quantum system. 
Specifying the dynamics amounts to a prescription for how to
obtain the density matrices on every edge from the density
matrix on the initial slice and the operators at the
vertices of the graph.  This prescription will be given
below in section~\ref{dyn}.

\section{Dynamics on Graphs}\label{dyn}
\subsection{Measurements and Interventions} We begin with
some standard material on density matrices and positive
operator-valued measures (POVMs)
\cite{Nielsen00,Preskill98}, before introducing Peres'
notion of
\emph{intervention operator} \cite{Peres00}.

Density matrices are used for describing quantum subsystems
which are part of larger quantum systems. 
In particular a local observer who has
access only to a subsystem $Q_1$ of a quantum system $Q$
will associate a density matrix to his subsystem.  Let
$\hi$ be the Hilbert space of state vectors of $Q$.

If the overall system $Q$ is in a state described by a wave
function
$\ket{\psi} \in \hi$, then its density matrix is the
operator $\rho =
\proj{\psi} \in End(\hi)$.  Since $Q$ can be decomposed into
subsystems, its Hilbert space is a tensor product $\hi =
\hi_{1} \otimes \hi_{2}$ of the Hilbert space $\hi_{1}$ of
the subsystem $Q_1$ and the Hilbert space
$\hi_{2}$ describing the remaining degrees of freedom.  The
density matrix of the subsystem $Q_1$ is then given by a
partial trace with respect the Hilbert space $\hi_{2}$:
$\rho_{1} = Tr^{\hi_2} \rho$.  If $\hi$ is any Hilbert
space, then the space of all density matrices will be denoted
$\mathsf{DM}(\hi)$.

The \emph{measurement} of a property of a quantum system
involves interaction with a classical apparatus.  When a
classical apparatus measures an observable of a quantum
subsystem sitting inside a larger system the appropriate
mathematical formalism for such generalized measurement is
that of \emph{positive operator-valued measure} or POVM. Let
the possible outcomes of the measurement be labelled by the
letter $\mu
\in \{1 \dots N\}$.  The measurement involves interaction
between the apparatus and the quantum system, described by a
unitary operator.  The classical apparatus has a preferred
basis of states indexed by $\mu$.  After the measurement,
the apparatus appears in one these preferred states.  Since
we are only interested in describing our quantum subsystem
$Q_1$, we trace out all the remaining degrees of freedom. 
Effectively to every outcome
$\mu$ is associated an operator $F_{\mu}$.  The density
matrix of $Q_1$ after the measurement with outcome $\mu$ is
given by
\begin{equation} 
\rho'_{\mu} = \frac{1}{p_{\mu}} F_{\mu} \rho
F^{\dagger}_{\mu} 
\end{equation} where $\rho$ is the density matrix before the
measurement and $p_{\mu}$ is a  numerical factor normalizing
the resulting density matrix to unit trace.   Consider the
family of positive operators $E_{\mu} = F^{\dagger}_{\mu}
F_{\mu}
$.  For a generalized measurement these have to satisfy the
condition 
$\sum_{\mu} E_{\mu} = I$.  The probability $p_{\mu}$ for
obtaining a  measurement result labelled by $\mu$ is then
given by: $p_{\mu} =  Tr E_{\mu} \rho$.  This justifies the
name POVM.  

Even more general measurement processes could be considered
if the observer  discards part of the quantum system during
the process of measurement.   The appropriate mathematical
formalism for describing these generalized  measurements is
that of {\it intervention operators} \cite{Peres00}.   In
the process of measurement, the density matrix changes
according to:
\begin{equation}\label{inap}
\rho'_{\mu} \ = \
\frac{1}{p_{\mu}} \sum_m A_{\mu m} \ \rho \ A_{\mu
m}^{\dagger}
\end{equation} The families of maps $A_{\mu m}$ now act in
general from one Hilbert space to another, i.e for fixed
$\mu$ and $m$ they correspond to rectangular matrices.

The label $\mu$ again distinguishes the set of possible
outcomes and the letter $m$ labels the degrees of freedom
discarded during this generalized measurement.  Since the maps
$A_{\mu m}$ come from measurements realized by unitary
operator on some larger Hilbert space they again satisfy a
completeness condition: $\sum_{\mu m} A^{\dagger}_{\mu m}
A_{\mu m} = I$, where $I$ is the identity operator in the
appropriate Hilbert space. Notice that if the labels $\mu$
and $m$ are absent in (\ref{inap}) the equation describes
unitary evolution.  Since the events we consider are
generalized measurements or unitary evolutions, equation
(\ref{inap}) is the appropriate mathematical representation
of those processes in full generality. Such maps
(\ref{inap}) on density matrices will be called {\it
intervention operators}.

\subsection{The dynamical prescription}\label{dyna}

We are now ready to start discussing the dynamics of a
quantum system represented by a dag $G$.  Dynamics will be
described by supposing that we are given a density matrix on
the initial spacelike slice, and then giving a prescription
for calculating the density matrices of future spacelike
slices. In essence, we are propagating the initial data
throughout the system. 

To each vertex $i\in G$ will be assigned an operator $T_i$,
and to each edge $e_j$ will be assigned a Hilbert space
$\hi_j$.  We note that all incoming (or outgoing) edges of a
given vertex are pairwise acausal and thus form a spacelike
slice.  Thus there will be a density matrix
$\rho_i^{in}$ associated to the slice of the incoming
edges.  Then one obtains the density matrix for the slice of
the outgoing edges by:
\[ \rho_i^{in}= T_i (\rho_i^{out}).\]

Notice that more generally, for two acausal vertices, the sets of
incoming or outgoing edges are pairwise acausal.  Thus, the
associated intervention operators will act on different
Hilbert spaces and hence commute.

\begin{figure}
\begin{center}
\epsfig{file=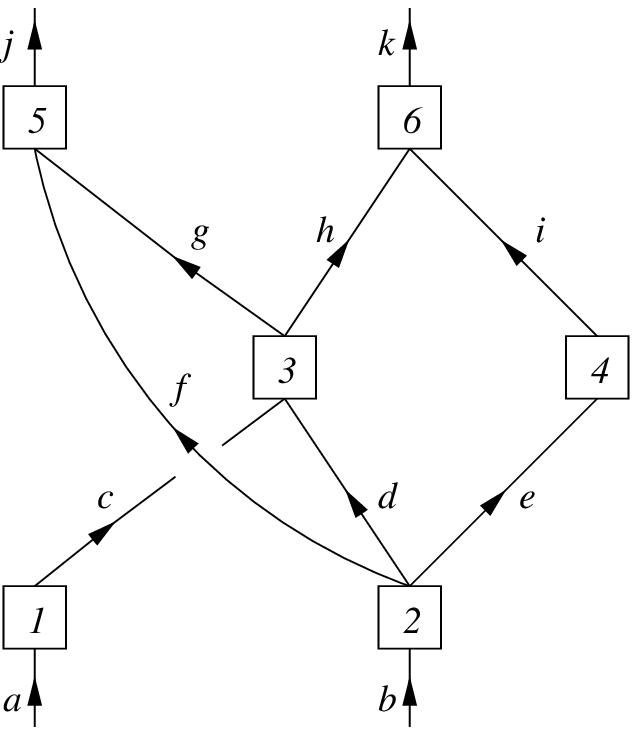}
\end{center}
\caption{}
\label{fig3}
\end{figure}

We begin with an illustrative example.  Consider  the dag of
Figure~\ref{fig3}.  
 Given the state on the initial slice, the
operators at the events propagate the state to the future. 
In the example of Figure~\ref{fig3} we have:
$\rho_c = T_1 (\rho_a)$,\ $\rho_{fde} = T_2 (\rho_b)$. 
However the next intervention operator $T_3$ must act on the
so far undefined density matrix
$\rho_{cd}$.  $T_3$ takes density matrices on
$\hi_c\ox\hi_d$ to those on
$\hi_g\ox\hi_h$.  By extending 
$T_3$ with the appropriate identity operators, we
can view it as a map from
$\mathsf{DM}(\hi_c\ox\hi_d\ox\hi_e\ox\hi_f)$ to
$\mathsf{DM}(\hi_e\ox\hi_f\ox\hi_g\ox\hi_h)$.  Then we can
define the density matrix on another space-like slice,
namely $\rho_{fghe} = T_3 (\rho_c \otimes \rho_{fde})$. 
Similarly $\rho_{fdi} = T_4 (\rho_{fde})$ and so on. 
Starting from density matrices on the initial edges and
using the intervention operators associated with the
vertices - extended with identities as needed - we obtain
density matrices on specific space-like slices.

The above inductive process for propagating density matrices
can be applied to any system described by a dag.  However,
the procedure only gives the density matrices for certain
spacelike slices within the dag.  For example, this
procedure does not yet yield a matrix for the slice $de$. 
To calculate such density matrices, we will also have to
make use of the trace operator.  Before extending the
procedure to such slices, we first consider those for which
the above process is sufficient.  We call these slices
\textit{locative}.
\begin{defin}{\rm Let $G$ be a dag, and $L$ a slice of $G$. 
Consider the set of all vertices $V$ which are to the past
of some edge in $L$.  Let $I$ be the set of initial edges in
the past of $L$.  Consider all paths of maximal length
beginning at an element of $I$ and only going through
vertices of $V$.  Then $L$ is \emph{locative} if all such
paths end with an edge in $L$.}
\end{defin} In our example, the locative slices are the
following:
$$a, b, ab, c, cb, def, adef, cdef, efgh, adfi, cdfi, fghe,
fghi, fgk, hej, hij, jk$$ while, for example, $de$ is not
locative.  Note that the fact that maximal slices are always
locative follows immediately from the definition of locative.

We now describe the general rule for calculating the density
matrices on locative slices.  Associated with each locative
slice $L$ is the set $I$ of initial edges in the past of
$L$.  We choose a family of slices that begins with $I$ and
ends with $L$ in the following way.  Consider the set of
vertices $V$ between the edges in $I$ and the edges in $L$. 
Because $L$ is locative we know that propagating slices
forwards through the vertices in
$V$ will reproduce $L$.  Let $M\subset V$ be such that the
vertices in $M$ are minimal in $V$ with respect to causal
ordering.  We choose arbitrarily any vertex $u$ in $M$,
remove the incoming edges of $u$ and add the outgoing edges
of $u$ to the set $I$ obtaining a new set of edges $I_1$. It
is clear that $I_1$ is spacelike and locative.  Proceeding
inductively in this fashion we obtain a sequence of slices
$I=I_0,I_1,I_2,\ldots,I_n = L$, where $n$ is the cardinality
of $V$.  Of course, this family of slices is far from unique.

The dynamics is obtained as follows.  Recall that the states
on initial edges are assumed not to be entangled with each
other so that one can obtain the density matrix on any set
of initial edges, in particular $I$, by a tensor product. 
Let $\rho_0$ be the density matrix on $I$.  We look at the
vertex $u$ that was used to go from $I$ to $I_1$ and apply
the intervention operator $T$ assigned to this vertex -
possibly augmented with identity operators as in the example
above.  Proceeding inductively along the family of slices,
we obtain the density matrix $\rho_n$ on $L$.

The important point now is that $\rho_n$ does not depend on
the choice of slicing used in going from $I$ to $L$.  This
can be argued as follows. Suppose we have a locative slice
$S$ and two vertices $u$ and $v$ which are both causally
minimal above $S$ and acausal with respect to each other.
Then we have four slices to consider, $S$, $S_u$, $S_v$ and
$S_{uv}$ where by $S_u$ we mean the slice obtained from $S$
by removing the incoming edges of $u$ and adding the
outgoing edges of $u$ to $S$ and similarly for the others. 
It is clear, in this case, that the intervention operators
assigned to $u$ and to $v$ commute and the density matrix
computed on
$S_{uv}$ is independent of whether we evolved along the
sequence
$S\to S_u\to S_{uv}$ or $S\to S_v\to S_{uv}$.  Now when we
constructed our slices at each stage we had the choice
between different minimal vertices to add to the slice.  But
such vertices are clearly pairwise acausal and hence, by the
previous argument applied inductively, the evolution
prescription is independent of all possible choices.

So far we have defined density matrices on locative slices
only.  To define density matrices on general spacelike
slices we will need to consider partial tracing operations.

\subsection{General Slices}

Recall that when one has subsystems $Q_1$ and $Q_2$ of a
quantum system
$Q$, the Hilbert space for $Q$ may be decomposed as
$\hi_1\ox\hi_2$ where
$\hi_i$ represents $Q_i$.  The density matrix for $Q_1$ is
obtained by tracing over $\hi_2$.  To obtain a candidate for
the density matrix of a spacelike slice $L$, we should find
a locative slice $M$ that contains $L$ and trace over the
Hilbert spaces on edges in $M\setminus L$.  Such a locative
slice $M$ always exists because maximal spacelike slices are
always locative.  $M$ is not unique however, and thus - as
we did for locative slices - we must show that different
choices give the same result.  To simplify the notation we
will discuss the case of density matrices associated with
single edges.  The case of a general space-like slice is
similar.

Consider an edge $e_i$ in a graph $G$.  Let $V_i =
\{v_{i_1}, \dots, v_{i_p}\}$ be the set of vertices in the
past of $e_i$.  Let $I_i =
\{e_{i_1}, \dots, e_{i_q}\}$ be the set of initial edges in
the past of
$e_i$.  Constructing a sequence of slices by incrementally
incorporating the vertices of $V_i$ in a manner similar to
what we did in the previous subsection, we get a locative
slice $M_i$ containing $e_i$.  Starting with the density
matrices on the edges of $I_i$ and applying the operators
associated with the vertices of $V_i$, we obtain the density
matrix on the locative slice $M_i$.  It is clear that $M_i$
is in an evident sense the minimal locative slice containing $e_i$. 

\begin{defin}{\rm We shall refer to $M_i$ as the \emph{least
locative slice} of the edge $e_i$.}  
\end{defin}   

Let the least locative slice $M_i$ of an edge $e_i$ consist
of edges $\{e_i, e_{j_1}, \dots, e_{j_r}\}$.  The density
matrix $\rho_{i,j_1,
\dots, j_r}$ on $M_i$ is an element of the space $\ind(\hi_i
\otimes
\hi_{j_1} \otimes \dots \otimes \hi_{j _r})$.  Let $Tr^{j_1
\dots j_r}$ be the partial trace operation $\ind(\hi_i
\otimes \hi_{j_1} \otimes \dots
\otimes \hi_{j_r}) \rightarrow \ind(\hi_i)$.
\begin{defin}[Density matrix associated with an
edge]\label{rho}\emph{The density matrix $\rho_i$ at the
edge $e_i$ is defined to be:
\begin{equation}
\rho_i \ = \ Tr^{j_1 \dots j_r} \ \rho_{i,j_1, \dots, j_r} .
\end{equation}}
\end{defin} If $M_i$ consists of the single edge $e_i$, then
no tracing is  done.

\begin{rem}
The causality condition for evolving the initial data on $G$
requires that the density matrix associated with a given
edge $e_i$ depends only on the initial data in the past of
$e_i$ and only those interventions to the past of $e_i$. 
The density matrix $\rho_i$ as defined in~\ref{rho}
satisfies this requirement by construction and so our
prescription for dynamical evolution is causal.
\end{rem}

In general, the edge $e_i$ is contained in many locative
slices and we could just as well have defined $\rho_i$ by
tracing over the complimentary degrees of freedom in any of 
these locative slices.  Independence of the resulting
density matrices  is the discrete analog of Lorenz (or
general) covariance in our framework.   To clarify the
discussion consider the quantum system represented by the
graph on Figure~\ref{figureF}.

\begin{figure}[htb]
\begin{center}
\setlength{\unitlength}{3947sp}%
\begingroup\makeatletter\ifx\SetFigFont\undefined%
\gdef\SetFigFont#1#2#3#4#5{%
  \reset@font\fontsize{#1}{#2pt}%
  \fontfamily{#3}\fontseries{#4}\fontshape{#5}%
  \selectfont}%
\fi\endgroup%
\begin{picture}(1974,2124)(4039,-4648)
\thinlines
\special{ps: gsave 0 0 0 setrgbcolor}\put(4801,-4261){\framebox(300,300){$1_{\hi_a}$}}
\special{ps: gsave 0 0 0 setrgbcolor}\put(4951,-4636){\vector( 0, 1){225}}
\special{ps: grestore}\special{ps: gsave 0 0 0 setrgbcolor}\put(4951,-4486){\line( 0, 1){225}}
\put(4741,-4486){$\rho_a$}
\special{ps: grestore}\special{ps: gsave 0 0 0 setrgbcolor}\put(5701,-3361){\framebox(300,300){$T$}}
\special{ps: gsave 0 0 0 setrgbcolor}\put(4951,-3961){\vector( 3, 2){450}}
\special{ps: grestore}\special{ps: gsave 0 0 0 setrgbcolor}\put(5401,-3661){\line( 3, 2){450}}
\put(5211,-3601){$\rho_c$}
\special{ps: grestore}\special{ps: gsave 0 0 0 setrgbcolor}\put(5851,-3061){\vector( 0, 1){300}}
\special{ps: grestore}\special{ps: gsave 0 0 0 setrgbcolor}\put(5851,-2836){\line( 0, 1){300}}
\put(5641,-2836){$\rho_d$}
\special{ps: grestore}\special{ps: gsave 0 0 0 setrgbcolor}\put(4951,-3961){\vector(-3, 4){450}}
\special{ps: grestore}\special{ps: gsave 0 0 0 setrgbcolor}\put(4501,-3361){\line(-3, 4){450}}
\put(4591,-3321){$\rho_b$}
\special{ps: grestore}\end{picture}
\end{center}
\caption{}
\label{figureF}
\end{figure}

Let the initial $\rho_a$ be the density matrix of a
maximally entangled state of two spin $1/2$ subsystems:
$\rho_a = |\psi_a\rangle\langle\psi_a|$, where $\psi_a =
1/\sqrt{2} \ (\psau \otimes \psbu + \psad \otimes \psbd)$. 
At the first vertex the two subsystems separate with no
classical intervention.  Therefore $\rho_{bc} = \rho_a$. 
The slice
$\{e_b, e_c\}$ is the least locative slice for the edge $e_b$
and we can compute the density matrix associated to this
edge: $\rho_b = T r^{c} \rho_{bc} = 1/2 \ (\psauk \psaub +
\psadk \psadb)$.  Next, let the intervention at the second
vertex be a measurement on the corresponding subsystem with
the result that the spin was found to be in the state
$\psbu$.  The intervention operator is the projection
operator on this state of the second subsystem: $T (\rho) =
2\ P_2^{\uparrow} \rho P_2^{\uparrow}$.  We obtain:
$\rho_{bd} = T (\rho_{bc}) = (\psauk \otimes \ \psbuk)
(\psaub \ \otimes
\psbub)$.  If now we attempt to trace $\rho_{bd}$ over the
subsystem associated with the edge $e_d$, we will obtain an
incorrect result for $\rho_b$, namely $\psauk \psaub$.   The
resolution is well known.  Since a classical observer
located on the edge $e_b$ is not aware of the result of the
intervention at the second vertex, for him the density
matrix $\rho_{bd}$ has evolved from $\rho_{bc}$ by an
operator $\tilde T$ which includes all possible outcomes of
the measurement:
$\tilde\rho_{bd} = {\tilde T} (\rho_{bc}) =
\sum_{s=\uparrow,\downarrow} P_2^s \rho_{bc} P_2^s$. 
Tracing out the
$d$-subsystem in the expression for $\tilde\rho_{bd}$, we
obtain the correct expression for $\rho_b$, namely  
$\rho_b = 1/2 \ (\psauk \psaub + \psadk \psadb)$.

Now we give the general prescription for computing the
density matrix on  an edge $e_i$ from an arbitrary locative
slice $L$ containing this edge.   We first compute a density
matrix $\tilde\rho_L$ for the slice $L$. But note this is
not the density matrix of definition~\ref{rho}. 
 
This density matrix is computed from the initial data by
applying intervention operators for the events in the past
of $L$ as before.  But now, we will consider two types of
events in the past of $L$, those that are in the past of
$e_i$ and those that are not.  For the events that are in
the past of the edge $e_i$, we use our regular intervention
operators without a summation over the set of possible
outcomes: $\rho \mapsto 1/p_{\mu} \sum_m A_{\mu m} \rho
A_{\mu m}^\dagger$.  We do not sum over the outcomes in this
case precisely because the outcome is in fact known at
$e_i$.  For the events that are in the past of the slice $L$
but not in the past of the edge $e_i$, we use operators
which sum over all possible outcomes: $\rho \mapsto
\sum_{\mu m} A_{\mu m} \rho A_{\mu m}^\dagger$. This time,
of course, the summation is there because the outcome cannot
be known at $e_i$ since these events are not in the past of
$e_i$.  

After we have obtained $\tilde\rho_L$, we trace out those
subsystems associated with edges in $L$ except for $e_i$ to
obtain the density matrix
$\tilde\rho_i$.  This is the density matrix associated with
our preferred edge $e_i$, as computed from the slice $L$. 
The independence of the result on the choice of $L$ is
expressed in the following proposition:

\begin{prop}[Covariance] Let $e_i$ be an edge in the dag
$G$.  The density matrix
$\rho_i$ associated  with the edge $e_i$ does not depend on
the choice of locative slice used to compute it.  
\end{prop}

\begin{proof} \\ We have already demonstrated that to any
edge $e_i$,  there is a unique least locative slice $M_i$
containing $e_i$.  Let
$\rho_i$ be the density matrix for the edge $e_i$ as
computed from the least locative slice and let
$\tilde\rho_i$ be the density matrix for the same edge but
computed from an arbitrary locative slice, say $L$,
containing $e_i$.  We will demonstrate the lemma by showing
that
$\rho_i=\tilde\rho_i$.

First note that $M_i$ being less than $L$ implies that there
is a set $V$ of events between $M_i$ and $L$.  The plan is
to remove the effect of these events and show that, at each
stage, the density matrix is unaffected.  We begin by
picking a maximal event, say $k$, with the intervention
operator
$T_k$.  Since $k$ is maximal and hence acausal with all
other maximal elements of $V$, as well as with all the
maximal elements to the past of
$e_i$, the intervention operator at $k$ commutes with all
the intervention operators at the vertices just mentioned. 
Thus, we can choose the intervention operator $T_k$ to be
the outermost, i.e.\ the density matrix
$\rho_L$ obtained by propagating to $L$ can be written as
\[ \rho_L = T_k(\rho') \] where $\rho'$ is the density
matrix on the (locative) slice obtained by removing the
edges to the future of $k$ from $L$ and adding the edges to
the past of $k$.  Using the explicit general form for an
intervention operator,
\[ \rho_L =
\sum_{\mu,m}A^{(k)}_{\mu,m}\rho'A^{\dagger(k)}_{\mu,m}.\] In
order to obtain the density matrix $\tilde\rho_i$, we trace
over all Hilbert spaces associated with edges in $L$ except
$e_i$.  In particular, we trace over the outgoing edges
associated with $k$.  Now we can use the cyclic property of
trace and rewrite the expression for $\tilde\rho_i$ as,
\[ \tilde\rho_i =
Tr(\sum_{\mu,m}A^{\dagger(k)}_{\mu,m}A^{(k)}_{\mu,m}\rho').
\] Now we use the identity
\[ \sum_{\mu m} A_{\mu m}^\dagger A_{\mu m}= I \] to get
\[ \tilde\rho_i = Tr(\rho').\]

We have eliminated the effect of the intervention operator at
$k$. Proceeding inductively we can peel off the intervention
operators associated with the rest of the vertices in $V$,
thus
\[ \tilde\rho_i = \rho_i.\]
\end{proof}

A similar argument for the case of a simple system
represented by the dag in Figure~\ref{figureD} is contained
in~\cite{Peres00b}.

\section{Polycategories}\label{poly}

We now wish to give a more axiomatic treatment of the above 
construction. This will require the use of several concepts from 
category theory and logic, which we now present.

We begin by introducing the algebraic or categorical concepts
necessary for our formulation of the dynamics of quantum
information flow.  While it might seem that these structures
are excessively abstract, this level of abstraction has
several advantages.  First, it provides a great deal of
generality.  Our definition can be applied in many contexts,
in particular it may be applied in situations other than the
sorts of information flow considered here.  Second, the two
crucial properties of interest, causality and covariance,
now become straightforward consequences of the functoriality
of our axioms.

\subsection{Posets, directed graphs and categories}

For comparison, we recall briefly that a poset is a set $P$
together with a binary relation on $P$ (i.e.  a subset of
$P\times P$) denoted $\leq$ that satisfies the properties of
antisymmetry, transitivity and reflexivity.  It is a natural
generalization of this idea to consider \emph{directed
graphs}.  A directed graph is simply a set $D$, the set of
\emph{vertices} or \emph{nodes}, together with a binary
relation $R$ on $D$.   No properties of $R$ are required in
the definition of directed graph.  In particular there is no
implicit transitivity assumed.  A directed graph has a
natural geometric visualization.  One considers the nodes as
points in the plane, and if $x$ and $y$ are nodes with
$\langle x,y\rangle\in R$,  we draw an arrow from $x$ to $y$.

As already remarked, the nodes of our directed graph will be
events, and arrows will represent propagation from one event
to another.  To avoid temporal loops, we will add the single
requirement that our directed graphs be \emph{acyclic},
i.e.  there does not exist a sequence of edges
$x_1,x_2,\ldots, x_n$ such that for all
$i\in\{1,2,\ldots,n-1\}$, we have
$\langle x_i,x_{i+1}\rangle\in R$,  and $x_1=x_n$.  This of
course corresponds to there being no directed cycles in the
geometric representation.  Hereafter, a directed acyclic
graph will be called a \emph{dag}.  Note that every poset,
considered as a directed graph, is acyclic.  This is a
consequence of transitivity and antisymmetry.  But dags are
a genuine generalization of posets.  

This difference will become more apparent when we consider
the space of
\emph{paths}.  In a poset all the paths are already included
(even if they are not explicitly drawn in the visualization
of the poset).  When we consider paths through a dag we may
have multiple paths between the same two vertices.  These
multiple paths represent different ways that information
flowed from one point to another, thus, we must regard them
as distinct.  Therefore - unlike the case with posets - we
do not just want to regard the resulting structure as a
binary relation, rather, we want to view it as a category.

It is natural to associate to any dag $D$, indeed to any
directed graph, a category.  We first briefly remind the
reader of the basic definitions. See \cite{Maclane98} for a
more extensive introduction.

\begin{defin} 
\rm{ A \textbf{category} \textsf{C} consists of two
collections, the collection of {\it objects} and the
collection of {\it morphisms}. Each morphism is assigned a
domain and codomain, both being objects of
\textsf{C}.  Typically we write $f\colon A\rarr B$ to mean
$f$ is a morphism with domain $A$ and codomain $B$.  To
every object $A$, we have a special morphism, the identity
$id\colon A\rarr A$.  There is also a composition law which
takes morphisms $f\colon A\rarr B$ and $g\colon B\rarr C$
and returns a morphism $gf\colon A\rarr C$.  All this data
must satisfy several evident equations, as described for
example in 
\cite{Maclane98}. We also remind the reader that a
\textbf{functor} is a morphism of categories, i.e.  a
functor, denoted $F\colon C\rarr D$ consists of a function
taking objects $c\in C$ to objects $F(c)\in D$ and taking
morphisms $f\colon c\rarr d$ in $C$ to morphisms $F(f)\colon
F(c)\rarr F(d)$.  A functor must preserve identities and
composition. } \end{defin}

To each dag $D$, we associate a category $\mathsf{C}(D)$. 
This is the category \emph{freely generated} by the dag. 
See for example \cite{Maclane98} Chapter 2, for a detailed
description.  The objects of our free category will be the
vertices of $D$.  If $x$ and $y$ are vertices, a morphism
from $x$ to $y$ is a directed path in our dag.  Identities are
paths of length 0, and composition is given by concatenation
of paths.  The verification of the axioms for a category is
straightforward.

One of the key points of our work is that we are proposing
passing from posets to categories.  As we have remarked
before, categories are more general than posets, indeed
posets correspond to a degenerate class of categories in
which there is at most one morphism between any two
objects.  The richer structure of categories allows us to
retain more information about the system.  Intuitively, the
use of categories allows us not merely to note that $x$
causally precedes
$y$, but to keep track of the different ways that $x$ may
evolve into $y$.   To make this more precise, we need
a slightly different construction on dags, which will 
yield polycategories as opposed to categories.  

\subsection{Polycategories}

Roughly speaking, the distinction between categories and
polycategories is the following: A category allows one to
have morphisms which go from single objects to single
objects.  A polycategory allows one to have morphisms from
lists of objects to lists of objects.  A typical morphism in
a polycategory (hereafter called a polymorphism) would be
denoted:

\[ f\colon A_1,A_2,\ldots,A_n\lrarr B_1,B_2,\ldots,B_m\]

There are a number of contexts in which such a generalization would be
useful.  Before giving the formal definition, we discuss two such contexts.
The first arises in algebra.  Consider Hilbert spaces, vector spaces or any
class of modules in which one can form a tensor product.  Then we can
define a polycategory as follows.  Our objects will be such spaces, and a
morphism of the above form will be a linear function:

\[ f\colon A_1\ox A_2\ox\ldots\ox A_n\lrarr B_1\ox
B_2\ox\ldots\ox B_m\]

Thus polycategories have proven to be quite useful in the analysis of
(ordinary) categories in which one can form tensor products of objects.
Indeed this was the original motivation for their definition.  See
\cite{Lambek69,Szabo75}.  Categories in which one has a reasonable notion
of tensor product are called \emph{monoidal}, and have recently figured
prominently in several areas of mathematical physics, most notably
topological quantum field theory \cite{Atiyah90,Baez95}.

The second well-known application of polycategories is to
logic.  Typically logicians are interested in the analysis of
\emph{sequents}, written:
\[ A_1,A_2,\ldots,A_n\vdash B_1,B_2,\ldots,B_m\]
\noindent Now 
$A_1,A_2,\ldots,A_n,B_1,B_2,\ldots,B_m$  represent formulas
in some logical system.  We say that the above sequent holds
if and only if the conjunction of $A_1,A_2,\ldots,A_n$
logically entails the disjunction of $B_1,B_2,\ldots,B_m$. 
There is a well-established correspondence between the sort
of logical entailments considered here and categorical
structures.  See for example
\cite{Lambek86}.

But notice the difference between this and our first
example.  When talking about vector spaces, the ``commas''
on the left and right were both interpreted as the tensor
product.  However in the logic example, we have two
different interpretations.  Commas on the left are treated
as conjunction, while commas on the right are treated as
disjunction.  Thus for a proper categorical interpretation
of polycategories, one needs categories with two monoidal
structures which interact in an appropriate fashion.  Such
categories are called \emph{linearly} or \emph{weakly
distributive}, a notion due to Cockett and Seely
\cite{Cockett97,Blute96}. Linearly distributive categories
are the appropriate framework for considering a specific
logical system known as \emph{linear logic}, introduced by
Girard \cite{Girard87,Girard89}.  For a brief exposition of
linear logic, see the appendix. As we will see, the refined
logical connectives of linear logic will be used to express
the entanglements of our system. 

There is a very geometric or graphical calculus for
representing morphisms in polycategories, which was
introduced by Joyal and Street in
\cite{Joyal91}, and given a logical interpretation in
\cite{Blute96}.  A polymorphism of the form:
\[ f\colon A_1,A_2,\ldots,A_n\lrarr B_1,B_2,\ldots,B_m\]
\noindent is represented as follows:

\setlength{\unitlength}{.6in}

\begin{picture}(6,3)(-2,0)
\put(1.2,1){\framebox(3,.6){$f$}}
\put(1.5,1){\vector(0,-1){.6}}
\put(1.9,1){\vector(0,-1){.6}}
\put(2.7,1){\vector(0,-1){.6}}
\put(3,.7){\ldots}
\put(2.2,.7){\ldots}
\put(3.5,1){\vector(0,-1){.6}}
\put(3.9,1){\vector(0,-1){.6}}
\put(1.5,2.2){\vector(0,-1){.6}}
\put(1.9,2.2){\vector(0,-1){.6}}
\put(2.7,2.2){\vector(0,-1){.6}}
\put(3,1.9){\ldots}
\put(2.2,1.9){\ldots}
\put(3.5,2.2){\vector(0,-1){.6}}
\put(3.9,2.2){\vector(0,-1){.6}}
\put(1.15,2.1){$A_1$}
\put(1.55,2.1){$A_2$}
\put(2.9,2.1){$A_{n-1}$}
\put(4.0,2.1){$A_n$}
\put(1.15,.5){$B_1$}
\put(1.55,.5){$B_2$}
\put(2.85,.5){$B_{m-1}$}
\put(4.0,.5){$B_m$}
\end{picture}

Thus the polymorphism is represented as a box, with the
incoming and outgoing arrows labelled by objects. 
Composition in polycategories then can be represented
pictorially in a very natural fashion.  Before giving a
general discussion of composition in a polycategory, we
illustrate this graphical representation.  Suppose we are
given two polymorphisms of the following form:

\begin{center}
$f\colon A_1,A_2,\ldots,A_n\lrarr B_1,B_2,\ldots,B_m,C$\\
$g\colon C,D_1,D_2,\ldots,D_k\lrarr E_1,E_2,\ldots,E_j$
\end{center}

Note the single object $C$ common to the codomain of $f$ and 
the domain of $g$.  Then under the definition of
polycategory,  we can compose these to get a morphism of 
form:

\[g\circ_Cf\colon A_1,A_2,\ldots,A_n,D_1,D_2,\ldots,D_k\lrarr B_1,B_2,
\ldots,B_m,E_1,E_2,\ldots,E_j\]

The object $C$ which ``disappears'' after composition is called the \emph{cut
object}, a terminology derived from logic.  Note that we subscript the
composition by the object being cut.  This composition would be represented
by the diagram on Figure~\ref{FigureComp}:

\begin{figure}[htb]
\begin{picture}(6,4)(-1,-1.5)
\put(1.2,1){\framebox(3,.6){$f$}}
\put(1.5,1){\vector(0,-1){2.12}}
\put(1.9,1){\vector(0,-1){2.12}}
\put(2.7,1){\vector(0,-1){2.12}}
\put(3,.7){\ldots}
\put(2.2,.7){\ldots}
\put(3.5,1){\vector(0,-1){2.12}}
\put(3.9,1){\line(0,-1){.6}}
\put(1.5,2.2){\vector(0,-1){.6}}
\put(1.9,2.2){\vector(0,-1){.6}}
\put(2.7,2.2){\vector(0,-1){.6}}
\put(3,1.9){\ldots}
\put(2.2,1.9){\ldots}
\put(3.5,2.2){\vector(0,-1){.6}}
\put(3.9,2.2){\vector(0,-1){.6}}

\put(3.6,-0.5){\framebox(3,.6){$g$}}
\put(3.9,-.5){\vector(0,-1){.6}}
\put(4.3,-.5){\vector(0,-1){.6}}
\put(5.1,-.5){\vector(0,-1){.6}}
\put(5.4,-.8){\ldots}
\put(4.6,-.8){\ldots}
\put(5.9,-.5){\vector(0,-1){.6}}
\put(6.3,-.5){\vector(0,-1){.6}}
\put(3.9,.7){\line(0,-1){.6}}
\put(3.65,.5){$C$}
\put(4.3,2.23){\vector(0,-1){2.13}}
\put(5.1,2.23){\vector(0,-1){2.13}}
\put(5.4,.4){\ldots}
\put(4.6,.4){\ldots}
\put(5.9,2.23){\vector(0,-1){2.13}}
\put(6.3,2.23){\vector(0,-1){2.13}}
\end{picture}
\caption{}
\label{FigureComp}
\end{figure}

We only label the segment corresponding to the cut object, for ease of
reading.  Thus composition in a polycategory is represented by the
concatenation of the graphs of $f$ and $g$, followed by joining the
incoming and outgoing edges corresponding to the cut object.  There are
several other possibilities for applications of the composition rule.  In
some cases, the graphical representation requires our arrows to cross.
This corresponds to having a \emph{symmetric} polycategory.  This is very
much related to having a symmetric tensor or tensors, i.e.  ones with the
property that $A\ox B\cong B\ox A$.  We will always assume our
polycategories are symmetric.

We now give a more formal definition of polycategory.  We
refer the reader to \cite{Cockett97,Szabo75} for further details.

\begin{defin}
\rm{
A \textbf{polycategory} $\mathsf{C}$ consists of
the following data:

\begin{itemize}
\item A set of objects, denoted $|\mathsf{C}|$.
\item If $A_1,A_2,\ldots,A_n$ and  $B_1,B_2,\ldots,B_m$ are
finite  sequences of  objects, then we have a set of
morphisms of the form 
$f\colon A_1,A_2,\ldots,A_n\lrarr B_1,B_2,\ldots,B_m$.  We note that
technically one must consider these sequences of objects as being defined
only up to permutation.
\item For every object $A$, we have an identity 
morphism $id_A\colon A\rarr A$.
\end{itemize}

\noindent The composition law was already described pictorially.  The data
of course are subject to a number of axioms, of which most important for us
is the one which requires associativity of composition.  The
notion of \emph{polyfunctor} between polycategories is also straightforward
to formulate.  One first has a function $F$ taking objects to objects, and
then given a morphism $f\colon A_1,A_2,\ldots,A_n \lrarr B_1,B_2, \ldots,
B_m$, one assigns to it a morphism
\begin{equation}
F(f)\colon F(A_1),F(A_2),\ldots, F(A_n) \lrarr F(B_1),F(B_2),\ldots,F(B_m).
\end{equation}
Again, a number of axioms must be satisfied, in particular the polyfunctor 
must commute with the composition of polymorphisms.  
}
\end{defin}

As suggested by the above, there is a relationship between polycategories 
and monoidal categories.  It is summarized in the following lemma, which 
can be found for example in \cite{Cockett97}:

\begin{prop}\label{thelemma}
Let $\mathsf{C}$ be a monoidal category.  Then one can associate to
$\mathsf{C}$ a polycategory (which will typically be denoted by
$P(\mathsf{C})$ as follows:
\begin{itemize}
\item The objects of $P(\mathsf{C})$ will be the same as those of 
$\mathsf{C}$.
\item A polymorphism of the form $f\colon A_1,A_2,\ldots,A_n\lrarr B_1,B_2,
\ldots,B_m$ is a morphism $f\colon A_1\ox A_2\ox\ldots\ox A_n\lrarr B_1\ox
B_2\ox\ldots\ox B_m$.  
\item Composition is induced by the composition in $\mathsf{C}$ in the
following way.  Suppose that we have two polymorphisms in $P(\mathsf{C})$
as follows:

\begin{center}
$f\colon A_1,A_2,\ldots,A_n\lrarr B_1,B_2,\ldots,B_m,C$\\
$g\colon C,D_1,D_2,\ldots,D_k\lrarr E_1,E_2,\ldots,E_j$
\end{center}

\noindent Then since we are in a monoidal category, we have morphisms
\begin{center}
$f\colon A_1\ox A_2\ox \ldots\ox A_n\lrarr B_1\ox B_2\ox\ldots\ox B_m\ox C$\\
$g\colon C\ox D_1\ox D_2\ox \ldots\ox D_k\lrarr E_1\ox E_2\ox \ldots\ox E_j$
\end{center}

The composite in  $P(\mathsf{C})$ is then given by:
\begin{equation}\label{identi}
g\circ_C f=(id_{B_1\ox B_2\ox\ldots\ox B_m}\ox g)\circ(f\ox id_{D_1\ox
D_2\ldots\ox D_k})
\end{equation}
\end{itemize}
\end{prop}

We note that the concepts of polycategory and monoidal category are not
equivalent.  To obtain an equivalence, one needs to replace monoidal
categories with the more general notion of linearly distributive category
mentioned above.

Now we will demonstrate that a dag generates a polycategory.  In this
construction, the nodes of the dag will be assigned morphisms and the edges
will be assigned objects.

We consider the dag example of Figure~\ref{figure6}.  We have changed
labels to be more appropriate for the present 
discussion.

\begin{figure}[htb]
\setlength{\unitlength}{3947sp}%
\begingroup\makeatletter\ifx\SetFigFont\undefined%
\gdef\SetFigFont#1#2#3#4#5{%
  \reset@font\fontsize{#1}{#2pt}%
  \fontfamily{#3}\fontseries{#4}\fontshape{#5}%
  \selectfont}%
\fi\endgroup%
\begin{picture}(1524,2124)(1800,-3223)
\thinlines
\special{ps: gsave 0 0 0
setrgbcolor}\put(4801,-1861){\framebox(300,300){$f_3$}}
\special{ps: gsave 0 0 0
setrgbcolor}\put(6001,-1861){\framebox(300,300){$f_4$}}
\special{ps: gsave 0 0 0
setrgbcolor}\put(4801,-2761){\framebox(300,300){$f_1$}}
\special{ps: gsave 0 0 0
setrgbcolor}\put(6001,-2761){\framebox(300,300){$f_2$}}
\special{ps: gsave 0 0 0
setrgbcolor}\put(4951,-2461){\vector( 0, 1){375}}
\special{ps: grestore}\special{ps: gsave 0 0 0
setrgbcolor}\put(4951,-2161){\line( 0, 1){300}}
\put(4731,-2161){$C$} 
\special{ps: grestore}\special{ps: gsave 0 0 0
setrgbcolor}\put(6151,-2461){\vector( 0, 1){375}}
\special{ps: grestore}\special{ps: gsave 0 0 0
setrgbcolor}\put(6151,-2161){\line( 0, 1){300}}
\put(6221,-2161){$E$}
\special{ps: grestore}\special{ps: gsave 0 0 0
setrgbcolor}\put(4951,-2461){\vector( 2, 1){600}}
\special{ps: grestore}\special{ps: gsave 0 0 0
setrgbcolor}\put(5551,-2161){\line( 2, 1){600}}
\put(5451,-2061){$D$}
\special{ps: grestore}\special{ps: gsave 0 0 0
setrgbcolor}\put(4951,-1561){\vector( 0, 1){375}}
\special{ps: grestore}\special{ps: gsave 0 0 0
setrgbcolor}\put(4951,-1186){\line( 0, 1){ 75}}
\put(4731,-1256){$F$}
\special{ps: grestore}\special{ps: gsave 0 0 0
setrgbcolor}\put(6151,-1561){\vector( 0, 1){375}}
\special{ps: grestore}\special{ps: gsave 0 0 0
setrgbcolor}\put(6151,-1186){\line( 0, 1){ 75}}
\put(6221,-1256){$G$}
\special{ps: grestore}\special{ps: gsave 0 0 0
setrgbcolor}\put(4951,-3211){\vector( 0, 1){225}}
\special{ps: grestore}\special{ps: gsave 0 0 0
setrgbcolor}\put(4951,-3061){\line( 0, 1){300}}
\put(4731,-3061){$A$}
\special{ps: grestore}\special{ps: gsave 0 0 0
setrgbcolor}\put(6151,-3211){\vector( 0, 1){225}}
\special{ps: grestore}\special{ps: gsave 0 0 0
setrgbcolor}\put(6151,-3061){\line( 0, 1){300}}
\put(6221,-3061){$F$}
\special{ps: grestore}
\end{picture}

\setlength{\unitlength}{3947sp}%
\begingroup\makeatletter\ifx\SetFigFont\undefined%
\gdef\SetFigFont#1#2#3#4#5{%
  \reset@font\fontsize{#1}{#2pt}%
  \fontfamily{#3}\fontseries{#4}\fontshape{#5}%
  \selectfont}%
\fi\endgroup%
\caption{}
\label{figure6}
\end{figure}

The idea behind the construction is that the nodes of the
dag (the  boxes in our picture)  will correspond to
polymorphisms.  For example, in the above picture,  the box
$f_1$ determines a polymorphism:

\[f_1\colon A\lrarr C,D\]

Similarly, $f_4$ determines a polymorphism $f_4\colon
D,E\rarr G$.  Thus we see that one has a polymorphism
corresponding to each node.  The domain of that polymorphism
will be the labels of the incoming arrows, and the codomain
is determined by the labels of the outgoing arrows.  These
are the basic morphisms of the polycategory.  As in the
previous construction, one must adjoin morphisms
corresponding to the allowable compositions.  For example,
in the above case, we can compose the morphisms $f_4$ and
$f_1$ along the cut object $D$ to obtain a new polymorphism
$f_4\circ_D f_1\colon A,E \rarr C,G$.  One must also add
identities and must force these composites to satisfy the
appropriate equations.  This construction yields the {\it
polycategory freely generated by the dag}.  More generally,
we would have the following definition.

\begin{defin}\label{free}{\rm We suppose that we are given a
finite dag $G$. The {\em free polycategory generated by
$G$}, denoted $P(G)$, is defined as follows.  If a given
vertex $v$ has incoming edges
$A_1,A_2,\ldots, A_n$ and outgoing edges $B_1,B_2,\ldots,B_m$
then the polycategory will have a polymorphism of the form
$f_v\colon A_1,A_2,\ldots,A_n\rarr B_1,B_2,\ldots,B_m$. In
general by induction,  if $P(G)$ has polymorphisms of the
form:

\begin{center}
$f\colon A_1,A_2,\ldots,A_n\lrarr B_1,B_2,\ldots,B_m,C$\\
$g\colon C,D_1,D_2,\ldots,D_k\lrarr E_1,E_2,\ldots,E_j$
\end{center}

\noindent then we require the existence of a composite
$g\circ_C f$ as a new polymorphism. We assume the existence
of an identity  morphism for each edge of $G$. Finally we
impose on this data the necessary equations implied by the
definition of polycategory. }
\end{defin}

\subsection{Categories of interventions} Next we describe an
appropriate for our construction polycategory of intervention
operators; there are several reasonable choices, this being 
the most straightforward. We start with the well known fact that  the
category $\mathsf{Hilb}$ of  Hilbert spaces and bounded
linear operators is a monoidal category.   Hence by the
construction of lemma \ref{thelemma}, we obtain a
polycategory.  However this is not the category we will
ultimately use. We will introduce a category
$\mathsf{Conj}$. Intuitively, the objects are Hilbert space 
endomorphisms and morphisms are conjugations. A more formal
definition is as  follows. Objects are finite-dimensional
Hilbert spaces. A morphism from
$\hi_1$ to $\hi_2$ is a finite family of maps $\{A_i\}_{i\in
I}$ of linear morphisms $A_i\colon \hi_1\rarr\hi_2$.
Composition is then described as  follows. If we have the
following pair of maps:

\[ \hi_1\to^{\{A_i\}_{i\in I}}\hi_2
\to^{\{B_j\}_{j\in J}}\hi_3\]

\noindent then the composite is:

\[ \hi_1\to^{\{B_j\circ A_i\}_{\langle i,j\rangle\in I\times J}}\hi_3\]

A morphism in $\mathsf{Conj}$ can be seen as taking
endomorphisms of
$\hi_1$ to endomorphisms of $\hi_2$ by the formula
$\cO\mapsto\sum_m A_m {\cal O} A_m^{\dagger}$. The monoidal
structure on $\mathsf{Hilb}$ lifts to a monoidal structure
on the category $\mathsf{Conj}$.  The tensor product
operator is the usual tensor product of operators on Hilbert
spaces, on maps we take all possible pairings.  We next
restrict the class of morphisms by considering only those
families suxh that the corresponding conjugation is trace
preserving. We call
the resulting category $\mathsf{Dio}$.  This category also
inherits a monoidal structure.  As discussed in Lemma
\ref{thelemma} any monoidal category canonically gives rise
to a polycategory associated to it.  We will denote by
$\cP(\mathsf{Dio})$ the polycategory associated with
$\mathsf{Dio}$.
      
\section{The logic of polycategories}\label{logic}

While definition~\ref{free} gives the free polycategory
generated by  a dag $G$, it will prove to be useful to have
a more constructive description.    Proof-theoretic
techniques have proven to be useful in describing free polycategories. 
In our case, the logical structures necessary are quite
simple, and so we  digress briefly to put definition~\ref{free} in 
logical terms.  Recall
that one of the common interpretations of a polymorphism is
as a logical sequent\footnote{We note that for
purposes of this paper sequents should always be considered
"up to permutation", i.e. one may rearrange the order of
premises and conclusions as one sees fit.} of the form:

\[ A_1,A_2,\ldots,A_n\vdash B_1,B_2,\ldots,B_m\]

Our system will have only one inference rule, called 
the \textit{Cut rule}, which states:
\begin{center}
\mbox{
\infer{\Gamma,\Gamma'\vdash\Delta,\Delta'}{\Gamma\vdash\Delta, A &
\Gamma',A\vdash\Delta'}}
\end{center}

This should be interpreted as saying that if one has derived
the two sequents above the line, then one can infer the
sequent below the line. Proofs in the system always begin
with \textit{axioms}.  Axioms are of the form 
$A_1,A_2,\ldots,A_n\vdash B_1,B_2,\ldots,B_m$, where
$A_1,A_2,\ldots, A_n$ are the incoming edges of some vertex
in our dag, and
$B_1,B_2,\ldots,B_m$ will be the outgoing edges. There will
be one such  axiom for each vertex in our dag. For example,
consider Figure~\ref{fig3}. Then we will have the following
axioms:
\[
 a\stackrel{1}{\ent} c\;\;\;
 b\stackrel{2}{\ent}  d,e,f\;\;\;
 c,d\stackrel{3}{\ent}  g,h\;\;\;
 e\stackrel{4}{\ent}  i\;\;\;
 f,g\stackrel{5}{\ent}  j\;\;\;
 h,i\stackrel{6}{\ent}  k
\] where we have labelled each entailment symbol with the
name of the corresponding vertex. The following is an
example of a deduction in this  system of the sequent
$a,b\vdash f,g,h,i$.

\begin{center}
\mbox{
\infer{a,b\vdash f,g,h,i}
{\infer{a,b\vdash e,f,g,h}{b\vdash 
d,e,f & \infer{a,d\vdash g,h}{a\vdash c & c,d\vdash g,h}}
& e\vdash i}
}
\end{center}
This deduction corresponds to the fact that in the free
polycategory generated by this dag, one has a morphism
$a,b\rarr f,g,h,i$. In fact, it is easy to see that there is
a precise correspondence between deductions in  this logical
system and nonidentity morphisms in the free polycategory. 

As a first attempt at capturing quantum evolution on a dag $G$ 
axiomatically, one
might consider taking a polyfunctor from $P(G)$ to
$P(\mathsf{Hilb})$, where
$\mathsf{Hilb}$ is the usual category of finite-dimensional
Hilbert spaces with its usual tensor product. Note that such
a polyfunctor must necessarily take a sequence of, say, incoming edges
$A_1,A_2,\ldots,A_n$ to
$\hi_1\ox\hi_2\ldots\ox\hi_n$ where $\hi_i$ corresponds to
$A_i$. Then one would (tentatively) define a set
$\Delta$ of edges to be {\it valid} if there is a
deduction in the logic generated by $G$ of
$\Gamma\vdash\Delta$ where
$\Gamma$ is a set of initial edges. Equivalently there must
be a morphism
$\Gamma\rarr\Delta$ in $P(G)$. Then the polyfunctor would
take this to a morphism of Hilbert spaces
$T\colon\hi_\Gamma\rarr\hi_\Delta$. The initial density
matrices would always be assumed to be given, and one would
just apply $T$ to the appropriate initial density matrices
to obtain the density matrix associated to $\Delta$. The locative 
slices are the ones on which density matrices can be obtained without 
the trace operation and we are looking to equate the notions of 
locative and valid for slices. This
approach would be genuinely axiomatic, and would evidently
be applicable to other situations by simply using a category
other than Hilbert spaces as the target of the polyfunctor.
Furthermore we would suggest that using logic as the means
of calculating the matrices gives the approach a very
canonical flavor.

However, with this notion of validity, we would fail to
capture all locative slices, and thus our tentative notion
of validity will have to be modified.  For example, consider
the dag underlying the system of Figure~\ref{figureD} shown
in Figure~\ref{figure5}.

\begin{figure}[htb]
\begin{center}
\setlength{\unitlength}{3947sp}%
\begingroup\makeatletter\ifx\SetFigFont\undefined%
\gdef\SetFigFont#1#2#3#4#5{%
  \reset@font\fontsize{#1}{#2pt}%
  \fontfamily{#3}\fontseries{#4}\fontshape{#5}%
  \selectfont}%
\fi\endgroup%
\begin{picture}(2724,3849)(3589,-6073)
\thinlines
\special{ps: gsave 0 0 0 setrgbcolor}
\put(4801,-3061){\framebox(300,300){$4$}}
\special{ps: gsave 0 0 0 setrgbcolor}
\put(3601,-4261){\framebox(300,300){$2$}}
\special{ps: gsave 0 0 0 setrgbcolor}
\put(6001,-4261){\framebox(300,300){$3$}}
\special{ps: gsave 0 0 0 setrgbcolor}
\put(4801,-5461){\framebox(300,300){$1$}}
\special{ps: gsave 0 0 0 setrgbcolor}\put(3751,-3961){\vector( 4, 3){600}}
\special{ps: grestore}\special{ps: gsave 0 0 0 setrgbcolor}\put(4351,-3511){\line( 4, 3){600}}
\put(4441,-3611){$d$}
\special{ps: grestore}\special{ps: gsave 0 0 0 setrgbcolor}\put(6151,-3961){\vector(-4, 3){600}}
\special{ps: grestore}\special{ps: gsave 0 0 0 setrgbcolor}\put(5551,-3511){\line(-4, 3){600}}
\put(5351,-3611){$e$}
\special{ps: grestore}\special{ps: gsave 0 0 0 setrgbcolor}\put(4951,-5161){\vector(-4, 3){600}}
\special{ps: grestore}\special{ps: gsave 0 0 0 setrgbcolor}\put(4351,-4711){\line(-4, 3){600}}
\put(4441,-4651){$b$}
\special{ps: grestore}\special{ps: gsave 0 0 0 setrgbcolor}\put(4951,-2761){\vector( 0, 1){450}}
\special{ps: grestore}\special{ps: gsave 0 0 0 setrgbcolor}\put(4951,-2386){\line( 0, 1){150}}
\put(4701,-2416){$f$}
\special{ps: grestore}\special{ps: gsave 0 0 0 setrgbcolor}\put(4951,-5886){\vector( 0, 1){125}}
\special{ps: grestore}\special{ps: gsave 0 0 0 setrgbcolor}\put(4951,-5836){\line( 0, 1){375}}
\put(4701,-5836){$a$}
\special{ps: grestore}\special{ps: gsave 0 0 0 setrgbcolor}\put(4951,-5886){\line( 0,-1){ 75}}
\special{ps: grestore}\special{ps: gsave 0 0 0 setrgbcolor}\put(4951,-5161){\vector( 4, 3){600}}
\special{ps: grestore}\special{ps: gsave 0 0 0 setrgbcolor}\put(5551,-4711){\line( 4, 3){600}}
\put(5351,-4651){$c$}
\special{ps: grestore}\end{picture}
\end{center}
\caption{}
\label{figure5}
\end{figure}
 
Corresponding to this dag, we get the following basic
morphisms (axioms):
\[a\vdash b,c \,\,\,\,\,\,\, b\vdash d\,\,\,\,\,\,\, c\vdash
e\,\,\,\,\,\,\, d,e\vdash f.\] Evidently, the set $\{f\}$ is
a locative slice, and yet  the sequent $a\vdash f$ is not
derivable. The sequent $a\vdash d,e$ is derivable, and one
would like to cut it against $d,e\vdash f$, but one is only
allowed to cut a single formula. Such ``multicuts'' are
expressly forbidden, as they lead to undesirable logical
properties
\cite{Blute93}.

Physically, the reason for this problem is that the sequent
$d,e\vdash f$ does not encode the information that the two
states at $d$ and $e$ are correlated.  It is precisely the
fact that they are correlated that implies that one would
need to use a multicut. To avoid this problem, one must
introduce some notation, specifically a syntax 
for specifying such correlations. We will
use the {\it logical connectives} of the multiplicative
fragment of {\it linear logic} \cite{Girard87,Girard95}
to this end. The multiplicative disjunction of linear
logic, denoted $\girpar$ and called the {\it par} connective, 
will express such nonlocal correlations.
In our example, we will write the sequent corresponding to
vertex $4$ as
$d\girpar e\vdash f$ to express the fact that the subsystems associated 
with these two edges are possibly entangled through interactions in their 
common past. 

Note that whenever two (or more)
subsystems emerge from an interaction, they are correlated. 
In linear logic, this is reflected by the following rule
called the (right) \emph{Par rule}:

\begin{center}
\mbox{
\infer[]{\Gamma\vdash \Delta,A\girpar
B}{\Gamma\vdash \Delta,A, B} }
\end{center}
Thus we can always introduce the symbol for correlation in the right 
hand side of the sequent.   

Notice that we can cut along a compound formula without
violating any logical rules.  So in the present
setting, we would have the following deduction:
\begin{center}
\mbox{
\infer{a\vdash f}
   {\infer{a\vdash d\girpar e}{\infer{a\vdash d, e}
{\infer{a\vdash c,d}{a\vdash b,c & b\vdash d}  & c\vdash e}}
& d\girpar e\vdash f} }
\end{center} All the cuts in this deduction are legitimate;
instead of a multicut we are cutting along a compound
formula in the last step. So the first step in modifying our
general prescription is to extend our polycategory logic, which originally
contained only the cut rule, to
include the connective rules of linear logic. These are described in 
the appendix. 

The above logical rule determines how one introduces a par connective
on the righthand side of a sequent. For the lefthand side, 
one introduces pars in the axioms by the following general prescription.
Given a vertex in a
multigraph, we suppose that it has incoming edges
$a_1,a_2,\ldots,a_n$ and outgoing edges $b_1,b_2,\ldots,b_m$.
In the previous formulation, this vertex would have been
labelled with the axiom $\Gamma=a_1,a_2,\ldots,a_n\vdash
b_1,b_2,\ldots,b_m$. We will now introduce several pars
($\girpar$) on the lefthand side to indicate entanglements
of the sort described above. Begin by defining a relation
$\sim$ by saying $a_i\sim a_j$ if there is an initial edge
$c$ and directed paths from $c$ to $a_i$ and from $c$ to
$a_j$. This is not an equivalence relation, but one takes
the equivalence relation generated by the relation $\sim$. 
Call this new relation $\cong$. This equivalence relation,
like all equivalence relations, partitions the set $\Gamma$
into a set of equivalence classes. One then "pars" together
the elements of each equivalence class, and this determines
the structure of the lefthand side of our axiom. For
example, consider vertices 5 and 6 in Figure~\ref{fig3}.
Vertex 5 would be labelled by $f\girpar g\vdash j$ and
vertex 6 would be labelled by $h\girpar i\vdash k$. On the
other hand, vertex 3 would be labelled by $c,d\vdash g,h$.

Just as the par connective indicates the existence of past 
correlations, we use the more familiar tensor symbol $\ox$, 
which is also a connective of linear logic, to indicate the lack of 
nonlocal correlation. This connective also has a logical rule:

\begin{center}
\mbox{
\infer{\Gamma,\Gamma'\vdash \Delta,\Delta',A\ox B}
{\Gamma\vdash \Delta,A & \Gamma'\vdash \Delta',B} }
\end{center}   

But we note that unlike in ordinary logic, this rule can only 
be applied in situations that are physically meaningful. We will say
that two deductions $\pi$ and $\pi'$ are {\it spacelike
separated} if all the the vertices of $\pi$ and $\pi'$ 
are pairwise spacelike separated. In the above formula, we
require that the deductions of $\Gamma\vdash \Delta,A$ and 
$\Gamma'\vdash \Delta',B$ are spacelike separated.
This restriction of application of
inference rules is similar to the restrictions of
{\it ludics} \cite{Girard01}. 
>From a categorical standpoint, the restrictions imply
that the connectives are only partial functors, but this is
only a minor issue. 

Summarizing, to every dag $G$ we associate its ``logic'', 
namely the edges are considered as formulas and vertices are 
axioms. We have the usual linear logical connective rules,
including the cut rule which in our setting is interpreted physically
as propagation. The par connective denotes correlation, and the tensor 
lack of correlation. Note that every deduction in our system
will conclude with a sequent of the form $\Gamma\vdash\Delta$,
where $\Gamma$ is a set of initial edges.

Now one would
like to modify the definition of validity to say that a set
of edges $\Delta$ is {\it valid} if in our extended
polycategory logic, one can derive a sequent
$\Gamma\vdash\hat{\Delta}$ such that the list of edges
appearing in $ \hat{\Delta}$ was precisely $\Delta$, and
$\Gamma$ is a set of initial edges. However this is still
not sufficient as an axiomatic approach to capturing all
locative slices. We note the example in Figure~\ref{figure8}.

\begin{figure}[htb]
\begin{center}
\setlength{\unitlength}{3947sp}%
\begingroup\makeatletter\ifx\SetFigFont\undefined%
\gdef\SetFigFont#1#2#3#4#5{%
  \reset@font\fontsize{#1}{#2pt}%
  \fontfamily{#3}\fontseries{#4}\fontshape{#5}%
  \selectfont}%
\fi\endgroup%
\begin{picture}(2124,3024)(3589,-5173)
\thinlines
\special{ps: gsave 0 0 0 setrgbcolor}\put(3601,-3061){\framebox(300,300){$f_3$}}
\special{ps: gsave 0 0 0 setrgbcolor}\put(3601,-4561){\framebox(300,300){$f_1$}}
\special{ps: gsave 0 0 0 setrgbcolor}\put(5401,-3061){\framebox(300,300){$f_4$}}
\special{ps: gsave 0 0 0 setrgbcolor}\put(5401,-4561){\framebox(300,300){$f_2$}}
\special{ps: gsave 0 0 0 setrgbcolor}\put(3751,-4261){\vector( 3, 2){675}}
\special{ps: grestore}\special{ps: gsave 0 0 0 setrgbcolor}\put(4426,-3811){\line( 3, 2){1125}}
\special{ps: grestore}\special{ps: gsave 0 0 0 setrgbcolor}\put(3751,-3061){\line( 3,-2){813.462}}
\special{ps: grestore}\special{ps: gsave 0 0 0 setrgbcolor}\put(5539,-4278){\vector(-3, 2){813.462}}
\put(4226,-3336){$d$}
\special{ps: grestore}\special{ps: gsave 0 0 0 setrgbcolor}\put(4726,-3736){\line( 1,-1){ 75}}
\put(4926,-3336){$h$}
\special{ps: grestore}\special{ps: gsave 0 0 0 setrgbcolor}\put(5551,-4261){\vector( 0, 1){675}}
\special{ps: grestore}\special{ps: gsave 0 0 0 setrgbcolor}\put(5551,-3661){\line( 0, 1){600}}
\put(5751,-3661){$e$}
\special{ps: grestore}\special{ps: gsave 0 0 0 setrgbcolor}\put(3751,-4261){\vector( 0, 1){675}}
\special{ps: grestore}\special{ps: gsave 0 0 0 setrgbcolor}\put(3751,-3661){\line( 0, 1){600}}
\put(3551,-3661){$c$}
\special{ps: grestore}\special{ps: gsave 0 0 0 setrgbcolor}\put(3751,-2761){\vector( 0, 1){450}}
\special{ps: grestore}\special{ps: gsave 0 0 0 setrgbcolor}\put(3751,-2386){\line( 0, 1){225}}
\put(3551,-2386){$f$}
\special{ps: grestore}\special{ps: gsave 0 0 0 setrgbcolor}\put(5551,-2761){\vector( 0, 1){450}}
\special{ps: grestore}\special{ps: gsave 0 0 0 setrgbcolor}\put(5551,-2386){\line( 0, 1){225}}
\put(5751,-2386){$g$}
\special{ps: grestore}\special{ps: gsave 0 0 0 setrgbcolor}\put(3751,-5161){\vector( 0, 1){300}}
\special{ps: grestore}\special{ps: gsave 0 0 0 setrgbcolor}\put(3751,-4936){\line( 0, 1){375}}
\put(3551,-4936){$a$}
\special{ps: grestore}\special{ps: gsave 0 0 0 setrgbcolor}\put(5551,-5161){\vector( 0, 1){300}}
\special{ps: grestore}\special{ps: gsave 0 0 0 setrgbcolor}\put(5551,-4936){\line( 0, 1){375}}
\put(5751,-4936){$b$}
\special{ps: grestore}\end{picture}
\end{center}
\caption{}
\label{figure8}
\end{figure}

Evidently the slice $\{f,g\}$ is locative, but we claim that
it cannot be derived even in our extended logic. To this
directed graph, we would associate the following axioms:

\[ a\vdash c,h \,\,\,\,\,\, b\vdash d,e\,\,\,\,\,\,
c,d\vdash f\,\,\,\,\,\, h,e\vdash g\]

Note that there are no correlations between $c$ and $d$ or
between $h$ and $e$. Thus no $\girpar$-combinations can be introduced. 
Now if one attempts to derive $a,b\vdash f,g$, we proceed as
follows:

\begin{center}
\mbox{
\infer{a,b\vdash h,e,f}{\infer{a,b\vdash c\ox d,h,e}{a\vdash
c,h & b\vdash d,e} & \infer{c\ox d\vdash f}{c,d\vdash f}}
}\end{center}

At this point, we are unable to proceed. Had we attempted
the symmetric approach tensoring $h$ and $e$ together, we
would have encountered the same problem. 

The problem is that our logical system is still missing one
crucial aspect, and that is that correlations develop
dynamically as the system evolves, or equivalently as the 
deduction proceeds. Thus our axioms must
change dynamically as well. We give the following definition.

\begin{defin}{\em
Suppose we have a deduction $\pi$ of the sequent
$\Gamma\vdash\Delta$ in the graph logic associated to the
dag $G$, and that $T$ is a vertex in $G$ to the future or acausal 
to the edges of the set
$\Delta$ with $a$ and $b$ among the incoming edges of $T$.
Then $a$ and $b$ are {\em correlated} with respect to $\pi$
if there exist outgoing edges $c$ and $d$ of the proof 
$\pi$ and directed paths from $c$ to $a$ and from $d$ to $b$.
}\end{defin}

So the point here is that when performing a deduction, one does 
not assign an axiom to a given vertex until it is necessary to use 
that axiom in the proof. Then one assigns that axiom using this new 
notion of correlation and the equivalence relation defined above. 
This prescription reflects the physical reality that entanglement of 
local quantum subsystems could develop as a result of a distant interaction 
between some other subsystems of the same quantum system. 
We are finally able to give the following crucial definition:

\begin{defin}{\em
A set
$\Delta$ of edges in a dag $G$ is said to be {\em valid} if there is a
deduction in the logic associated to $G$ of
$\Gamma\vdash\hat{\Delta}$ where $\hat{\Delta}$ is a sequence of formulas
whose underlying set of edges is precisely $\Delta$ and where
$\Gamma$ is a set of initial edges, in fact the set of initial edges
to the past of $\Delta$.}
\end{defin}

We are also ready to state the result relating the logical deduction and 
the dynamics of Section~\ref{dyna} in a graph. 

\begin{thm} A set of edges is valid if and only if it is locative. More 
specifically, if there is a deduction of $\Gamma\vdash\hat{\Delta}$
as described above, then $\Delta$ is necessarily locative. Conversely,
given any locative slice, one can find such a deduction.
\end{thm}
\begin{proof}\\ Recall that a locative slice $L$ is obtained from the set
of initial edges in its past by an inductive procedure.  
At each step, we choose arbitrarily a minimal vertex $u$ in the past of $L$,
remove the incoming edges of $u$ and add the outgoing edges. This step 
corresponds to the application of a cut rule, and the method we have used
of assigning the par connective 
to the lefthand side of an axiom ensures that it is 
always a legal cut. The tensor rule is necessary in order to combine 
spacelike separated subsystems in order to prepare for the application of 
the cut rule.
\end{proof}

Thus we have successfully given an axiomatic logic-based approach
to describing evolution. In summary, to find the density matrix 
associated to a locative slice $\Delta$, one finds a set 
of linear logic formulas whose underlying set of atoms is $\Delta$ 
and a deduction of $\Gamma\vdash\hat{\Delta}$ where $\Gamma$ is 
as above. This deduction is interpreted as a morphism in the 
corresponding polycategory, and the polyfunctor to $\cP(\mathsf{Dio})$
is applied to obtain a morphism in the category $\mathsf{Dio}$. 
(Note that in this context a polyfunctor is furthermore required to take 
any tensor or par connective in $\Gamma$ or $\hat{\Delta}$ to the 
usual tensor in $\mathsf{Dio}$.)
One then plugs in the given initial data to obtain the density matrix
corresponding to that slice. Given a nonlocative slice, one simply finds
a locative slice containing it, repeats the above procedure and then 
traces out the extraneous edges.

\section{Conclusions}\label{conc}

We have presented an axiomatic system for the analysis 
of quantum evolution. The dynamics is local as to preserve causality, but at 
the same time entanglement of separated quantum systems is faithfully 
represented. One could apply these ideas 
to other situations
by using a category other than the category of intervention
operators
as the target of the functor. An appropriate
categorical structure for the target is the notion of  
a {\it traced monoidal category} \cite{Joyal96} or
the notion of a {\it traced ideal} \cite{Blute99}. 
See also \cite{Blute00}.
One particular situation which might be analyzed in this
framework is the notion of classical probabilistic
information. The paper \cite{Blute99} contains a category of
{\it probabilistic relations} which might be of particular
interest in this setting.

Our work also suggests a natural
extension of the notion of {\it consistent} or {\it
decoherent histories} \cite{Gell-Mann93,Griffiths96}.
Restricting the intervention operators at the vertices of our graph $G$ 
to be projection operators we can consider $G$ to denote a particular 
history within a set of histories. This relaxes the usual linear ordering of 
events considered in the literature thus far.
An exposition of histories on graphs is under preparation. 

\section*{Acknowledgements} The authors would like to thank
NSERC for its financial support.  We would also like to
thank Phil Scott and Jean-Yves Girard for inviting us to
present this work at the joint SMF-AMS conference in Lyon.  The
paper
\cite{Markopoulou00}, which led to our initial consideration
of these ideas, was pointed out to us by Ioannis Raptis.  We
would like to especially thank Rafael Sorkin for a lengthy
discussion on causal sets and related topics.  Finally, the
second author would like to thank the University of Ottawa
Department of Mathematics and Statistics for its support. 
\bibliography{causal}

\appendix
\section{Linear logic}

This section can safely be skipped by logicians.

Linear logic~\cite{Girard87} is a logic introduced by Girard
in 1987 to allow a finer analysis of how ``resources'' are
consumed in the course of a deduction. As already remarked
in the text, the primary objects of study in logic and
especially proof theory are {\it sequents}, and the
constructors of sequents, the {\it inference rules}. Several
examples have already been given such as the {\it cut rule}:

\begin{center}
\mbox{
\infer[CUT]{\Gamma,\Gamma'\vdash\Delta,\Delta'}{\Gamma\vdash\Delta,
A &
\Gamma',A\vdash\Delta'}}
\end{center}

So typically an inference rule is a prescription for
creating a more complex sequent from one or possibly several 
simpler ones. Two typical inference rules are the rules of
{\it contraction} and {\it weakening}. These are as follows:

\begin{center}
\mbox{\infer[CONT]{\Gamma,A\vdash\Delta}
{\Gamma,A,A\vdash\Delta}}
\end{center}

\begin{center}
\mbox{\infer[WEAK]{\Gamma,A\vdash\Delta}
{\Gamma\vdash\Delta}}
\end{center}

There are similar rules for the righthand side as well. These
have long been standard in most logics, and indeed have a
strong intuitive meaning. For example, contraction says that
it is unnecessary to make the same assumption twice.
However, in Girard's reexamination of the sequent calculus,
he proposed an interpretation in which the formulas to the
left of a sequent are resources to be consumed in the course
of producing the output, i.e. the conclusions. From this
perspective, the rules of contraction and weakening are
quite dubious. The first step towards defining linear logic
then is to recover these rules from the system. The result is
a remarkably rich structure, the most notable aspect of
which is that the usual connectives of logic, conjunction
and disjunction, each split into two connectives. These
connectives are naturally split into two classes, the {\it
multiplicative} and the {\it additive} connectives. It is
only the multiplicative connectives that will concern us
here. Here are the rules for these connectives:

\begin{center}
\mbox{
\infer[Right-\girpar]{\Gamma\vdash \Delta,A\girpar
B}{\Gamma\vdash \Delta,A, B} }
\end{center}   

\begin{center}
\mbox{
\infer[Left-\girpar]{\Gamma,\Gamma' A\wp B\vdash \Delta,\Delta'}
{\Gamma,A\vdash \Delta & \Gamma',B\vdash \Delta'} }
\end{center}   

\begin{center}
\mbox{
\infer[Right-\ox]{\Gamma,\Gamma'\vdash \Delta,\Delta',A\ox B}
{\Gamma\vdash \Delta,A & \Gamma'\vdash \Delta',B} }
\end{center}   

\begin{center}
\mbox{
\infer[Left-\ox]{\Gamma,A\ox B\vdash \Delta}{\Gamma,A,B\vdash \Delta} }
\end{center}

Categorically, the structure of linear logic has striking
properties as well. As is traditional in categorical logic,
one can form a category whose objects are formulas, and
morphisms are proofs. This construction is described for
example in \cite{Lambek69,Lambek86}. When one applies this
construction to (multiplicative) linear logic ({\bf MLL}),
one obtains a special class of symmetric monoidal closed
categories called {\it $*$-autonomous}. These were defined
by Barr in \cite{Barr79}. 

Subsequently it was demonstrated that the correspondence
between proofs in {\bf MLL} and morphisms in the free
$*$-autonomous category is quite sharp.  See
\cite{Blute93,Blute96}.   This correspondence between
morphisms and proofs is best expressed using {\it proof
nets}, a graph-theoretic system for representing {\bf MLL}
proofs \cite{Girard87}. Proof nets had already been seen to
be a remarkable deductive system, exhibiting properties of
great importance in the analysis of computation, especially
concurrent computation.  The precise connection between
proof nets and free $*$-autonomous categories
provides further evidence of their great utility.

\end{document}